\documentclass[a4paper,fleqn]{cas-dc}

\usepackage{natbib}
\usepackage{comment}
\usepackage[british]{babel}

\newcommand*{\id}{{\normalfont\hbox{1\kern-0.15em \vrule width .8pt depth-.5pt}}}

\begin{document}
\emergencystretch 3em

\def\floatpagepagefraction{1}
\def\textpagefraction{.001}
\shorttitle{FQHE in semiconductor systems}
\shortauthors{Z. Papi\'c and Ajit C. Balram}

\title [mode = title]{Fractional quantum Hall effect in semiconductor systems}

\author[1]{Zlatko Papi\'c}[orcid=0000-0002-8451-2235]
\cormark[1]
\ead{z.papic@leeds.ac.uk}

\address[1]{School of Physics and Astronomy, University of Leeds, Leeds LS2 9JT, United Kingdom}

\author[2,3]{Ajit C. Balram}[orcid=0000-0002-8087-6015]

\address[2]{Institute of Mathematical Sciences, CIT Campus, Chennai 600113, India}
\address[3]{Homi Bhabha National Institute, Training School Complex, Anushaktinagar, Mumbai 400094, India}

\cortext[cor1]{Corresponding author}

\begin{abstract}
The fractional quantum Hall (FQH) effect refers to the strongly-correlated phenomena and the associated quantum phases of matter realized in a two-dimensional gas of electrons placed in a large perpendicular magnetic field. In such systems, topology and quantum mechanics conspire to give rise to exotic physics that manifests via robust quantization of the Hall resistance. In this chapter, we provide an overview of the experimental phenomenology of the FQH effect in GaAs-based semiconductor materials and present its theoretical interpretations in terms of trial wave functions, composite fermion quasiparticles, and enigmatic non-Abelian states. We also highlight some recent developments, including the parton theory and the Dirac composite fermion field theory of FQH states, the role of anisotropy and geometrical degrees of freedom, and quantum entanglement in FQH fluids.

\end{abstract}


\maketitle

\section{Key objectives}

\noindent $\bullet$ describe the rich phenomenology exhibited by two-dimensional electrons hosted in a semiconductor system placed in a perpendicular magnetic field;
    
\noindent  $\bullet$ overview of different theoretical approaches for describing fractional quantum Hall (FQH) phases of matter;

\noindent  $\bullet$ introduce the concept of non-Abelian FQH states and their potential applications in fault-tolerant topological quantum computation;

\noindent  $\bullet$  summarize recent developments in the field, including parton theory of the FQH effect, geometric properties of FQH states, entanglement-based and field-theoretic approaches to understanding FQH systems.

\section{Notations and acronyms}

\noindent 
2DEG: two-dimensional electron gas \\
IQHE: integer quantum Hall effect \\
FQHE: fractional quantum Hall effect \\
LL: Landau level \\
LLL: lowest Landau level \\
SLL: second Landau level \\
PH: particle-hole  \\ 
CF: composite fermion  \\
MR: Moore-Read \\
HLR: Halperin-Lee-Read \\
$\nu$: filling factor 
\\
$B$: magnetic field strength \\
$\ell{=}\sqrt{\hbar /eB}$: magnetic length \\
$z{=}x{-}iy$: complex number representation of the electron coordinates $(x,y)$ in the 2D plane \\
$\Phi_{n}$: wave function of $n$ filled Landau levels \\
$\Phi_1 {\equiv} \prod_{i{<}j}(z_i{-}z_j)$: Laughlin-Jastrow factor \\
$\mathcal{P}_{\rm LLL}$: projection to the lowest Landau level \\

\section{Introduction}
\label{sec:intro}

The fractional quantum Hall (FQH) effect occurs in a system of electrons confined to two spatial dimensions and a strong magnetic field perpendicular to the two-dimensional (2D) plane. In such systems, the kinetic energy is completely quenched by the magnetic field and Coulomb interaction exerts non-perturbative effects on the behavior of the electrons,  typically without analogs in systems of weakly-interacting electrons encountered elsewhere in condensed matter physics. In this chapter, we do not aim to present a historical overview of the FQH effect, but rather highlight the key theoretical concepts for its understanding and to illustrate the connections between a variety of different approaches that have been deployed to tackle the problem. Perhaps most surprisingly, first-quantized wave functions have turned out to be highly valuable tools for understanding FQH phenomena, in contrast to the second-quantized field theories widely used in other areas of many-body physics. To emphasize this unique aspect of the FQH effect, our presentation will focus on introducing physics from the perspective of trial wave functions, motivated by experimental observations.

Before we dive into its fascinating experimental phenomenology, it is instructive to present a high-level description of the core of FQH physics. The following ``Theory of Everything" Hamiltonian governing a FQH system should be familiar to most physicists: 
\begin{eqnarray}
\notag 	H & =& \sum_{j} \frac{1}{2 m_{\mathrm{b}}}\left[\boldsymbol{p}_{j}+\frac{e}{c} \boldsymbol{A}\left(\boldsymbol{r}_{j}\right)\right]^{2}+ \sum_{j} U\left(\boldsymbol{r}_{j}\right) 
+g \mu_{\mathrm{B}} \boldsymbol{B} \cdot \boldsymbol{S}  \\
&+& \frac{e^{2}}{4\pi\epsilon} \sum_{j<k} \frac{1}{\left|\boldsymbol{r}_{j}-\boldsymbol{r}_{k}\right|}.
\end{eqnarray}
The first line contains the one-body terms in the problem:  the kinetic energy in the presence of a constant external magnetic field $\boldsymbol{B}{=}\nabla{\times} \boldsymbol{A}$;  the potential $U$  incorporating the lattice, disorder and confinement effects; finally, the Zeeman energy. The second line is the Coulomb interaction energy between (non-relativistic) electrons. 

As a first step, let us simplify the problem down to its bare essentials. While disorder, confinement and lattice effects can be occasionally very important, we will neglect them here and set $U{=}0$. Most FQH experiments discussed below are performed using GaAs-AlGaAs heterostructures, in which electrons have a band mass of $m_{\mathrm{b}}{=}0.067 m_{\mathrm{e}}$ (where $m_{\mathrm{e}}$ is the electron mass in vacuum), dielectric constant of $\epsilon{=}12.6\epsilon_0$, and Landé $g$ factor $g{=}{-}0.44$. From this, we can estimate the energy required to flip a spin -- the Zeeman splitting -- to be $E_{\mathrm{Z}} {=} g \mu_{\mathrm{B}} B {\approx} {0.3} B[\mathrm{T}] \mathrm{K}$, where the magnetic field $B[\mathrm{T}]$ is in units of Tesla. With the exception of Sec.~\ref{sec:multi}, throughout this chapter, we will assume the electron spin is fully polarized due to a large magnetic field. 

In order to diagonalize what's left of the Hamiltonian, the usual strategy is to first diagonalize the one-body part of $H$ which gives rise to discrete Landau levels (LLs) of a single electron in a magnetic field. The Landau solution yields two important scales in the problem: the magnetic length
\begin{eqnarray}
\ell=\left(\frac{\hbar}{e B}\right)^{1 / 2} \approx \frac{25}{\sqrt{B[\mathrm{T}]}} \mathrm{nm},
\end{eqnarray}
and the cyclotron energy splitting between LLs
\begin{eqnarray}
\hbar \omega_{\mathrm{c}}=\hbar \frac{e B}{m_{\mathrm{b}}} \approx 20 B[\mathrm{T}] \mathrm{K}.
\end{eqnarray}
On the other hand, the typical Coulomb energy is given by
\begin{eqnarray}
V_{\mathrm{C}} \equiv \frac{e^{2}}{4\pi\epsilon \ell} \approx 50 \sqrt{B[\mathrm{T}]} \mathrm{K}.
\end{eqnarray}
For the sake of convenience, throughout we shall use natural units where $\ell{=}1$ and $e^{2}/ (4\pi \epsilon \ell){=}1$. 

To render the many-body problem as tractable as possible, one often makes a further approximation by choosing $V_\mathrm{C} {\ll} \hbar \omega_{\mathrm{c}}$. In this limit, the Coulomb interaction is too weak to scatter particles across different LLs, i.e., entire dynamics is contained within a single LL and there is no ``LL mixing". This assumption is not always warranted and its validity should be tested either numerically or experimentally. Nevertheless, under this approximation, the FQH Hamiltonian takes a remarkably simple form
\begin{eqnarray}	\label{eq:FQHE_Ham_LLL}
H=\mathcal{P} \left(  \frac{e^{2}}{4\pi\epsilon} \sum_{j<k} \frac{1}{\left|\boldsymbol{r}_{j}-\boldsymbol{r}_{k}\right|} \right) \mathcal{P},
\end{eqnarray}
i.e., the Hamiltonian is just the Coulomb interaction which must be diagonalized with the restriction of remaining in a single LL. The latter restriction is formally implemented by the global operator $\mathcal{P}$ which projects out all states that have a component outside the given LL. For most of our discussion below, the LL in question will be the lowest LL (LLL), although we will also mention the second Landau level (SLL) and higher LLs.

The model in Eq.~\eqref{eq:FQHE_Ham_LLL} embodies the theoretical essence of the FQH effect: it is believed to capture universal physical properties of broad classes of correlated phases emerging in the regime of the FQH effect. However, despite its seemingly simple form, it is impossible to analytically diagonalize the Hamiltonian of Eq.~\eqref{eq:FQHE_Ham_LLL}. The difficulty stems from  $\mathcal{P}$ being a global operator defined on a many-electron Hilbert space whose dimension grows exponentially with the number of electrons. Specifically, two main challenges are: (i) Eq.~\eqref{eq:FQHE_Ham_LLL}, has no parameters whatsoever, including no-small parameters that one might use to attempt a perturbative approach; (ii) If the Coulomb interaction is turned off, one ends up with an exponentially large degeneracy of any LL, again precluding an easy solution. 

 Broadly speaking, this chapter will focus on the following questions: (i) what are the possible phases of matter that arise from Eq.~\eqref{eq:FQHE_Ham_LLL} as one varies the electron density (or, alternatively, the magnetic field)? (ii) what are the universal properties of such phases in the long-wavelength limit, e.g., the types of order and excitations that govern the low-energy physics? (iii) what are the fundamental principles that allow to understand and microscopically model such phases? 

\begin{figure*}
	\centering
	\includegraphics[width=0.92\linewidth]{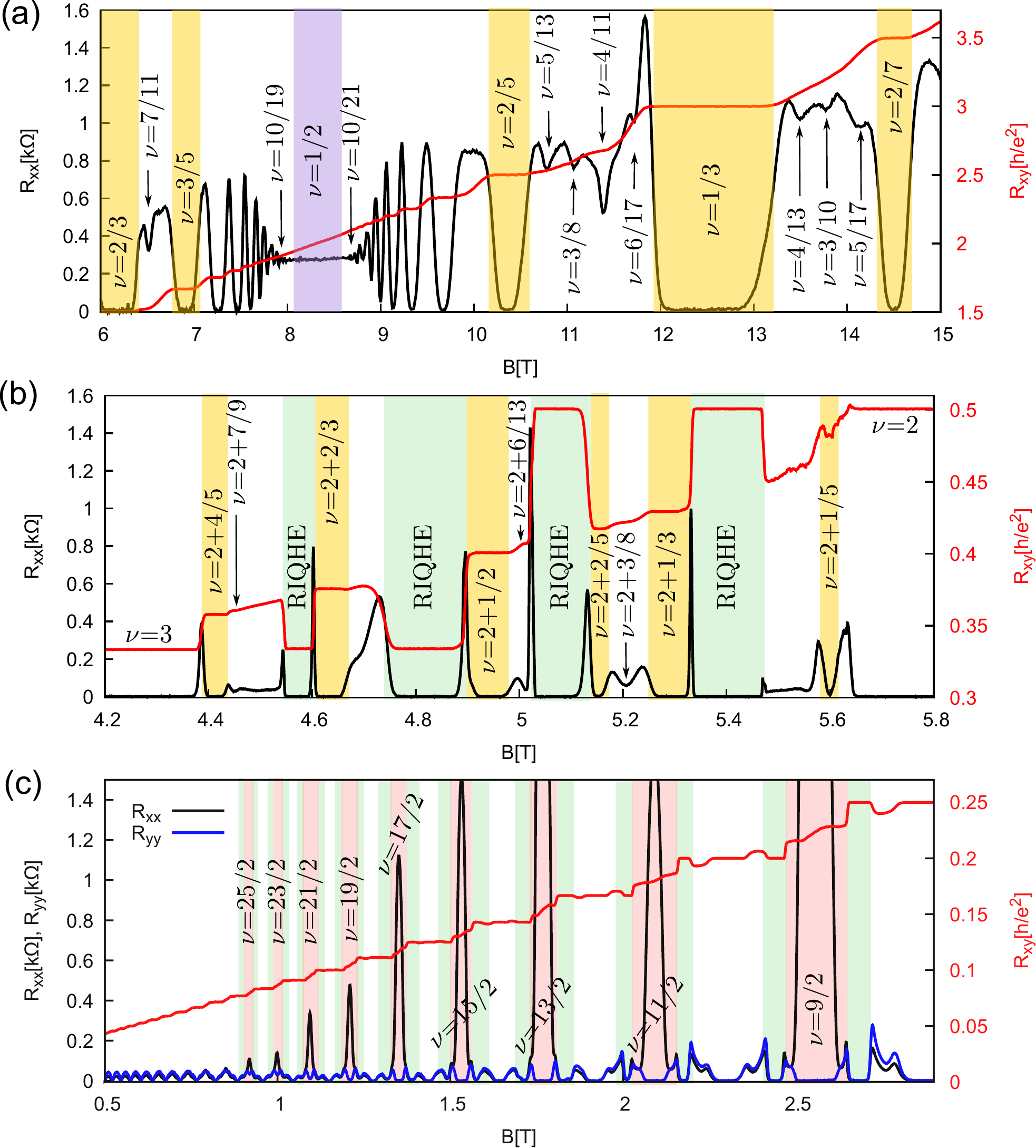}
	\caption{
	Experimental measurements of the Hall resistance $R_{xy}$ and the longitudinal resistance $R_{xx}$ as a function of the magnetic field $B$ applied perpendicularly to a 2DEG. 
	(a) Resistance traces in the lowest Landau level from Ref.~\cite{Pan03}. Prominent gapped states are shaded in yellow. Other weaker states (without fully developed minimum in $R_{xx}$) are also indicated. The composite Fermi liquid is shaded purple. 
	(b) In the second Landau level, in addition to gapped states at odd-denominator filling factors, there are also even-denominator states at $\nu{=}2{+}1/2$ and $\nu{=}2{+}3/8$. Reentrant integer quantum Hall states are shaded in green. Data is reproduced from Ref.~\cite{Kumar10}.
	(c) In higher Landau levels, the FQH  plateaus are replaced by states with  strongly anisotropic transport, as revealed by the large difference between $R_{xx}$ and $R_{yy}$. Examples are shaded in red and identified with stripe phases. Reentrant integer quantum Hall states are shaded in green. Data is taken from Ref.~\cite{Ro19}.
	}
	\label{fig:fqhetraces}
\end{figure*}

\section{Phenomenology of the FQH effect}
\label{sec: fqh_pheno}

In a typical FQH experiment, one measures the Hall resistance, extracted from the voltage drop perpendicular to the current flowing through a two-dimensional electron gas (2DEG) placed in a perpendicular magnetic field. The 2DEG should have sufficiently low concentration of impurities and high mobility of the electrons. The fabrication of such samples has required major technological advances that we will not touch upon. As found by Tsui, Stormer and Gossard in 1982~\cite{Tsui82}, upon sweeping the magnetic field,
the Hall resistance of such a 2DEG exhibits quantization at
\begin{eqnarray}\label{eq:Hallresistance}
R_{xy}=\frac{h}{\nu e^{2}},
\end{eqnarray}
where $\nu$ was originally found to take the value $1/3$. Subsequently, quantization was observed at other rational values of $\nu$ such as 2/3, 2/5, 3/7, etc. On the other hand, $\nu$ is related to the electron density $\rho$ and is known as the \emph{filling factor} or \emph{filling fraction},
\begin{eqnarray}
\nu=\frac{\rho h}{e B}.
\end{eqnarray}
In an earlier work by von Klitzing, Dorda and Pepper~\cite{Klitzing80}, similar resistance quantization had been observed for integral values of $\nu$, i.e., in the regime of integer quantum Hall (IQH) effect. In both cases, the resistance was found to be quantized not only at the special value of $\nu$ but also in some neighborhood around the special value, which manifests as a resistance plateau in the experimental trace, as illustrated in the experimental data shown in Fig.~\ref{fig:fqhetraces}(a). Furthermore, when $R_{xy}$ is pinned to a plateau, the longitudinal resistance (i.e., the one measured along the direction of the current) exhibits an Arrhenius behavior $R_{xx}{\propto} \exp \left({-}\Delta/(2 k_{\mathrm{B}} T)\right)$, where the energy scale $\Delta$ is interpreted as a gap to excitations. $R_{xx}$ vanishes in the limit $T {\rightarrow} 0$, indicating dissipationless transport. A deep minimum in $R_{xx}$ is often seen as a precursor of a developing FQH state, however, the definitive observation of a FQH state requires a quantized plateau in $R_{xy}$.

In both IQH and FQH cases, the $R_{xy}$ quantization is remarkably accurate to about one part in a billion. This is a manifestation of quantum-mechanical coherence across a semiconductor material on mesoscopic length scales. Moreover, as seen from the appearance of the Planck constant, the quantization is not described by classical electrodynamics. The latter also predicts that the resistance should vary linearly with the magnetic field. One hint to a topological origin of the phenomenon is that the resistance is quantized, not the resistivity (they both have the same units in two dimensions). Moreover, the Hall quantization was found to be universal: it is independent of the sample type, geometry, and various materials parameters, such as the band mass of the electron or the dielectric constant of the background semiconductor, and is robust to sufficiently weak disorder.

\subsection{Incompressible fluids}
\label{sec: incompressible_fluids}

\begin{figure}[htb]
	\centering
	\includegraphics[width=\linewidth]{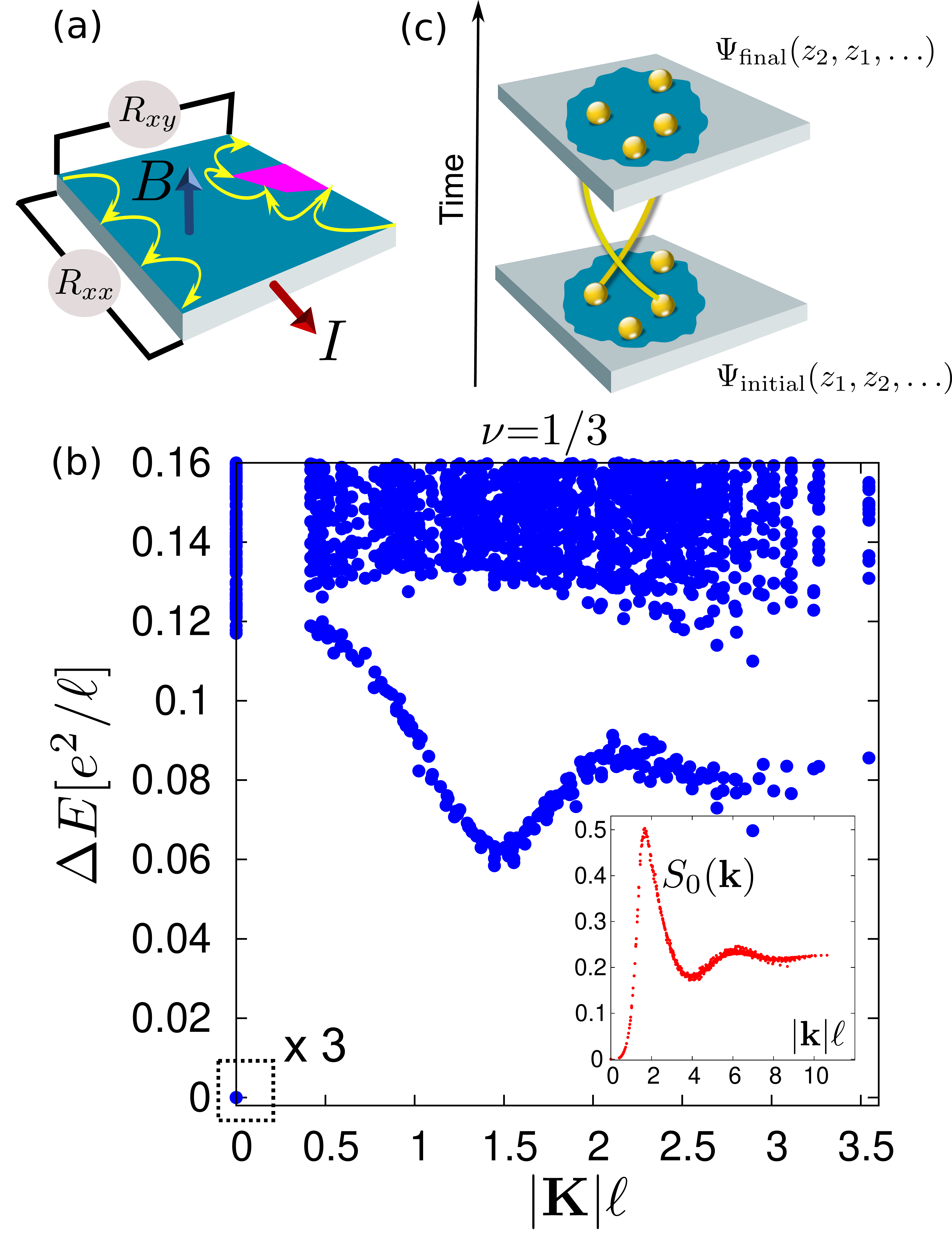}
	\caption{Incompressible fluids and topological order. (a) Transport experiment probing a FQH state. Robust quantization of the Hall resistance is associated with a current-carrying edge state (yellow), which is insensitive to a low concentration of impurities (depicted in purple). In a semiclassical picture, the edge state forms a narrow one-dimensional channel that propagates ballistically in one direction. (b) On a torus, the $\nu{=}1/3$ FQH ground state has a three-fold topological degeneracy and carries zero momentum $\mathbf{K}{=}0$, while the rest of the energy spectrum is separated by a gap. Low energy charge-neutral excitations form a collective mode. Data is obtained using DiagHam open source package~\cite{diagham} for exact diagonalization of the Coulomb interaction projected to the lowest LL for $N{\leq }12$ electrons. The same plot includes data for both square and hexagonal unit cells, illustrating the robustness of the low-energy physics to geometric details. Inset shows the projected guiding-center structure factor $S_0(\mathbf{k})$, which reveals a liquid ground state. (c) Elementary charged excitations of a FQH fluid behave as anyons: by taking two anyons around each other, the resulting wave function $\Psi_\mathrm{final}$ is not simply equal to ${\pm}\Psi_\mathrm{initial}$, as it would be for bosons or fermions. 
	}
	\label{fig:incompressible}
\end{figure}

Performing the transport measurements described above, one obtains remarkable traces of $R_{xy}$ and $R_{xx}$ as a function of the magnetic field $B$, as we illustrate in Fig.~\ref{fig:fqhetraces} using current state-of-the-art experimental data. At sufficiently low temperatures (typically, in mK range) and very high magnetic fields, the trace is dominated by FQH states which display a quantized plateau in $R_{xy}$ and vanishing $R_{xx}$. These are signatures of incompressible quantum fluids, which arise from strong Coulomb repulsion between the electrons residing in the partially-filled LLL, as the kinetic energy of the electrons is quenched. FQH quantum fluids are unconventional in that they represent topological phases of matter. While a detailed theory of such topological quantum fluids will be presented in subsequent sections, here we briefly summarize some of their defining  properties -- see Fig.~\ref{fig:incompressible}:

\begin{itemize}

\item FQH phases are quantum liquids that do not exhibit long-range order for any local order parameter. Thus, FQH phases cannot be described by a Landau-Ginzburg type theory based on spontaneous symmetry breaking~\cite{Prange87}. For example, the (LL-projected) structure factor of a FQH state, shown in inset of Fig.~\ref{fig:incompressible}(b), takes the form characteristic of a liquid phase, indicating that the FQH state is uniform in the bulk and distinct from, say, a Wigner crystal.

\item The bulk of a FQH fluid exhibits an energy gap $\Delta$, i.e., there is a discontinuity in the chemical potential as one sweeps the magnetic field. Furthermore, there is also a gap for charge-neutral excitations at the fixed magnetic field, as seen in Fig.~\ref{fig:incompressible}(b). Near the system's boundary, however, there are \emph{gapless} excitations which propagate chirally, in the direction determined by the external magnetic field, see Fig.~\ref{fig:incompressible}(a). In the semiclassical picture, these excitations can be visualized as ``skipping orbits"~\cite{Halperin82}, which can also go around weak impurity potentials, avoiding dissipation. 

\item  The `topological' nature of the FQH fluid derives from its sensitivity to the topology of the manifold it lives in. For example, in the infinite plane or when wrapped around a sphere, the fluid has a unique ground state, but when it is placed on a torus or any higher genus surface, the ground state develops a degeneracy~\cite{WenNiu}. At filling factor $\nu{=}p/q$, with $p$,$q$ co-prime positive integers, the ground state degeneracy is at least $q$~\cite{Haldane85b}. For example, in the $\nu{=}1/3$ state shown in Fig.~\ref{fig:incompressible}(b), the ground state is three-fold degenerate.

\item By threading magnetic flux or, equivalently, by twisting the torus boundary conditions, the degenerate ground states can mix with one another but remain separated from other excited states at energies $ {\geq} \Delta$. The total Hall conductance carried by the ground state multiplet is proportional to the topological invariant called the Chern number $C$~\cite{Thouless82}. This gives the Hall conductance for each member of the multiplet   $\sigma_{xy} {=} \nu \frac{e^2}{h}$. The topological nature of the Hall conductance explains its robustness at sufficiently low temperatures $k_B T {\ll} \Delta$.

\item Thermal Hall conductance of  FQH fluids is given by~\cite{Kane97}
    \begin{eqnarray}\label{eq:thermalHall}
     \kappa_{xy}{=}c_{-}[\pi^2 k^{2}_{B}/(3h)]T,
    \end{eqnarray}
where the universal coefficient $c_{-}$ is known as the chiral central charge. This expression holds at temperatures much smaller than the gap and under the assumption that edge modes are fully equillibrated. 

\item The excitations of incompressible FQH fluids 
are \emph{anyons}~\cite{Leinaas77, Wilczek82}: an exotic type of particle which carry a fraction of the electron charge and also have ``fractional" exchange statistics~\cite{Arovas84}. In particular, for states such as $\nu{=}1/3$ depicted in Fig.~\ref{fig:incompressible}(c), the adiabatic exchange of the coordinates of two anyons results in a complex phase, $\Psi_\mathrm{final}\left(z_{2}, z_{1}\right) = e^{i \theta} \Psi_\mathrm{initial}\left(z_{1}, z_{2}\right)$. Bosons and fermions are special cases with $\theta{=}0$ and $\theta{=}\pi$, respectively, while Abelian anyons correspond to other real values of $\theta$. Anyons with richer exchange statistics are also possible and will be discussed in Sec.~\ref{sec:nonabelian}. We note that recent experiments have observed statistical properties of the simplest types of Abelian anyons in the $\nu{=}1/3$ state using mesoscopic interferometry~\cite{Nakamura20} as well as anyon colliders~\cite{Bartolomei20}. 

\item FQH fluids can be described by an emergent Chern-Simons gauge field $\mathbf{a}$, which attaches magnetic fluxes to particles~\cite{Zhang89}. This enables the transmutation of statistics and the existence of anyon excitations. The Chern-Simons field contributes a topological term to the effective action of FQH fluids, 
\begin{eqnarray}
\mathcal{L}_\mathrm{CS}=\frac{1}{4\pi\nu}\epsilon^{\mu\nu\lambda}a_\mu \partial_\nu a_\lambda, 
\end{eqnarray}
where $\epsilon^{\mu\nu\lambda}$ is the Levi-Civita tensor and we assumed the FQH state is at filling $\nu{=}1/q$, with Einstein summation convention implied. Thus, the effective low-energy theory of FQH fluids is a variant of topological quantum field theory (see Sec.~\ref{sec:field} for further discussion).

\end{itemize}

The properties summmarized in this section are believed to be shared by all incompressible FQH states, which are primarily realized in the LLL and SLL, recall Fig.~\ref{fig:fqhetraces}(a) and (b). However, as we explain below, the microscopic mechanisms giving rise to these topological fluids can be very different. In particular, the FQH states in the LLL are markedly different from those in the SLL. In contrast to the majority of gapped FQH liquids presented in Fig.~\ref{fig:fqhetraces}(a), we note that there also exists a \emph{gapless} liquid state which is similar to the Fermi liquid state in an ordinary 2DEG in zero magnetic field. This state occurs near $\nu{=}1/2$ where one observes a smooth dependence of both the longitudinal and Hall resistances in Fig.~\ref{fig:fqhetraces}(a). Nevertheless, this Fermi-liquid-like state has many exotic properties of its own and it will be discussed separately in Sec.~\ref{sec: CFL}. 

We note here that FQH states extending on either side of $\nu{=}1/2$ are related by an operation that maps electrons to holes and vice versa, i.e., it relates states at $\nu$ with $1{-}\nu$. This particle-hole (PH)  symmetry is exact in the limit $B{\to}\infty$. In this limit, any two-body interaction, like the Coulomb one, results in eigenstates that are perfectly PH symmetric. We will return to the consequences of PH symmetry in Secs.~\ref{sec:apf} and~\ref{sec:field}.

\subsection{Broken symmetry phases}

In addition to incompressible liquids, other types of correlated phases of matter can occur in partially filled LLs, primarily in the regime of low filling factors or in higher LLs, see Fig.~\ref{fig:fqhetraces}(b)-(c). While interactions still underpin the formation of such phases, these phases spontaneously break the translation symmetry of the 2D plane by forming crystals or stripe patterns. Such phases are typically weakly-correlated and they have been successfully understood using the more traditional methods, such as Hartree-Fock theory.

\subsubsection{Wigner crystals in the lowest Landau level}
\label{sec: Wigner_crystals}

``Classical" electrons at sufficiently low densities in 2D form a triangular Wigner crystal (WC) to minimize the Coulomb repulsion between them. Strong magnetic field quenches the kinetic energy of the electrons, leaving only interactions to decide the fate of the system. This setting was believed to be ideal to stabilize the WC phase and provided one of the motivations for the Tsui-Stormer-Gossard experiment~\cite{Tsui82}. 

As it turned out, the FQH liquid has lower energy than WC at high filling factors~\cite{Yoshioka83}. However, at very low filling factors, when the density of electrons is sufficiently small, a WC is expected to be stabilized. Recent studies suggest that the FQH fluid can prevail over the WC at filling factors as low as $\nu{=}1/7$~\cite{Pan02, Chung22}. Furthermore, the WC that is stabilized at $\nu{<}1/7$ is not that of electrons, but of electron-vortex composites called composite fermions (CFs)~\cite{Archer13,Zuo20}, that we will introduce in Sec.~\ref{sec:CF_theory}.

While direct signatures of the Wigner solid phase have remained elusive, experiments have observed a number of indirect signatures. The presence of a WC has been probed by studying the commensurability oscillations of CFs in a nearby layer~\cite{Deng16}. Collective oscillations of the crystalline domains of the WC resulting from disorder give rise to ``pinning resonances" in the frequency-dependent conductivity which have been detected experimentally~\cite{Williams91, Zhu10}. The melting behavior of the magnetic-field-induced WC suggests that it harbors quantum correlations and is markedly different from the classical WC at zero-field~\cite{Ma20}. 

\begin{figure}[bth]
	\centering
	\includegraphics[width=0.95\linewidth]{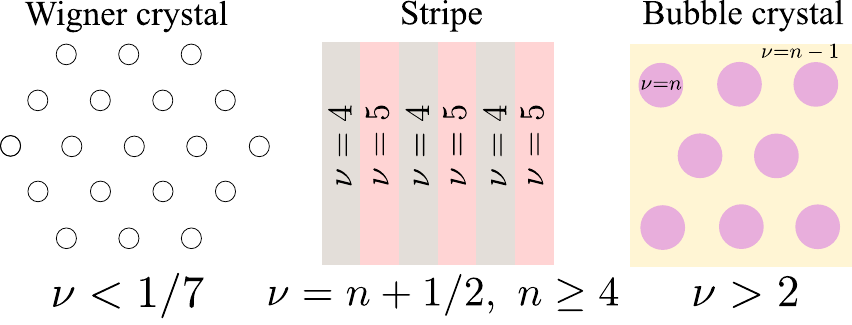}
	\caption{Some examples of symmetry-broken phases realized in a two-dimensional electron gas placed in a strong perpendicular magnetic field. 
	}
	\label{fig:phases_electronic_B}
\end{figure}

\subsubsection{Stripes, nematics and bubble crystals in higher LLs}
\label{sec: stripes_nematics_bubbles}

In higher LLs, interactions can also engender various phases that spontaneously break certain symmetries while exhibiting integral quantization of the Hall resistance, as seen  in Fig.~\ref{fig:fqhetraces}(b) and (c). In the vicinity of integral fillings, if disorder is sufficiently weak, the additional electrons or holes are expected to form a WC. Using Hartree-Fock theory, Koulakov, Fogler, and Shklovskii~\cite{Koulakov96, Fogler96} predicted that the extra electrons or holes in a high  LL have a tendency to form a WC, a ``bubble" phase -- a WC  of bubbles of electrons, where each bubble hosts multiple electrons,  or stripes of alternating fillings, depending on the filling factor. For example, when a high LL is nearly half full (i.e., the filling factor is $\nu{=}n{+}1/2$ with $n$ being a sufficiently large integer), strong transport anisotropy is observed, as shown in Fig.~\ref{fig:fqhetraces}(c), consistent with the formation of stripes~\cite{Du99, Lilly99}. The direction of the anisotropy aligns with one of the crystal axes, due to the explicit symmetry breaking via the host lattice. Moreover, the anisotropy can flip by $90^{\circ}$ as a function of the in-plane magnetic field~\cite{Lilly99} or the density~\cite{Zhu2002}. Many re-entrant transitions are seen in higher LLs~\cite{Eisenstein02, Xia04}, indicating competition between numerous approximately degenerate ground states~\cite{Shibata2001} with a common feature of integral Hall resistance quantization.

\section{Laughlin states}
\label{sec:Laughlin_fractions}

\subsection{Laughlin's wave function}
\label{sec:Laughlin_wf}

Soon after the experimental observation of the FQH plateau at $\nu{=}1/3$, Laughlin constructed a novel ansatz to describe the electronic ground state~\cite{Laughlin83}. Laughlin's wave function is given by:
\begin{equation}
 \Psi^{\rm Laughlin}_{\nu=1/(2p+1)} = \prod_{i<j}(z_{i}-z_{j})^{2p+1}
 \exp\left[-\sum_{k}\frac{|z_{k}|^{2}}{4\ell^2}\right] \equiv \Phi_{1}^{2p+1},
 \label{eq: Laughlin_gs}
\end{equation}
where $z_{k} = x_{k}{-}i y_{k}$ parametrizes the 2D coordinate of the $k^{\rm th}$ electron as a complex number, and $p$ is a positive integer ($p{=}1$ for $\nu{=}1/3$). The $\nu{=}1$ IQH wave function, $\Phi_{1}{=}\prod_{i<j}(z_{i}{-}z_{j})$, is referred to as the Laughlin-Jastrow factor. To simplify the notation, below we shall often drop the ubiquitous Gaussian factor and the subscript ``$\nu{=}$" from the wave functions.

The high power of the Jastrow factor in the Laughlin wave function builds additional correlations in the many-body state (beyond those mandated by the Pauli principle) which keeps the electrons away from each other and thus minimizes their Coulomb repulsion. Because of this, one may expect the wave function in Eq.~\eqref{eq: Laughlin_gs} to provide a good variational description of the exact ground state of the Hamiltonian in Eq.~\eqref{eq:FQHE_Ham_LLL}. Quite remarkably, it turned out that the overlap between the Laughlin wave function and the exact ground state of Eq.~\eqref{eq:FQHE_Ham_LLL} for as many as $N{=}15$ electrons is consistently above $98{\%}$, despite an exponential increase in the size of the Hilbert space with $N$ and the lack of any variational parameters in Eq.~\eqref{eq: Laughlin_gs}. This suggests that the Laughlin wave function captures all the essential correlations of the $\nu{=}1/3$ ground state even in the thermodynamic limit. Indeed, due to the ``nice" form of the Laughlin wave function, many of its properties such as energy, density, pair-correlation function etc. have been studied in large systems using the Monte Carlo method~\cite{Morf86b}. The (projected) structure factor of the $\nu{=}1/3$ state, shown in inset of Fig.~\ref{fig:incompressible}(b), confirms that the state describes an incompressible liquid.

Laughlin's theory furthermore predicts that the FQH effect can also occur at $\nu{=}1/5$ and a quantized Hall plateau at this filling was observed in experiments. Subsequent numerical simulations have shown that the Laughlin wave functions have high overlaps with the exact LLL Coulomb ground small systems at $\nu{=}1/3,~1/5$ and $1/7$~\cite{Fano86, Ambrumenil88}. For $\nu{<}1/7$, a Laughlin liquid is expected to give way to a crystalline state~\cite{Lam84}. By particle-hole conjugation in the LLL, Laughlin's theory also explains the plateaux observed at fractions $\nu{=}1{-}1/(2p{+}1)$ which can be understood as a $1/(2p{+}1)$ Laughlin state of ``holes".

\subsection{Charged excitations}\label{sec:charged}

Laughlin also constructed wave functions for the charged excitations in this system. Laughlin's wave function for the positively charged excitation, referred to as a \emph{quasihole} (qh), located at a position $\eta$, is given by:
\begin{equation}
    \Psi^{\rm Laughlin, qh}_{1/(2p+1)} = \prod_{i}(z_{i}-\eta)\Psi^{\rm Laughlin}_{1/(2p+1)},
    \label{eq: Laughlin_qh}
\end{equation}
while that of the negatively charged \emph{quasiparticle} (qp) is:
\begin{equation}
    \Psi^{\rm Laughlin, qp}_{1/(2p+1)} = \prod_{i}(2\partial_{z_{i}}-\bar{\eta})\Psi^{\rm Laughlin}_{1/(2p+1)}.
    \label{eq: Laughlin_qp}
\end{equation}
In Eq.~\eqref{eq: Laughlin_qp} the derivatives only act on the polynomial part of the wave function and not on the Gaussian factor. To motivate these wave functions, consider the situation where $\eta{=}0$. The polynomial factor $\prod_{i} z_{i}$ pushes electrons away from the origin (the wave function of Eq.~\eqref{eq: Laughlin_qh} vanishes at $\eta$) thereby depleting the electron density at the origin resulting in the creation of a positively charged quasihole. The derivatives, on the other hand, pull electrons in towards the origin which leads to a build-up of an excess charge at the origin resulting in the creation of a negatively charged quasiparticle. 

The probability density corresponding to Laughlin's ground state and quasihole wave functions can be interpreted as the partition function of a one-component classical plasma in 2D~\cite{LaughlinPrange}. Using this plasma analogy, Laughlin showed that the quasihole excitation carries a fractional charge of $e/(2p+1)$ (we denote the electronic charge by ${-}e$). Via PH conjugation, one can readily see that the quasiparticle excitation also carries a fractional charge of value $({-}e)/(2p{+}1)$. Shot-noise experiments have verified the presence of $e/3$ charged quasiparticles in the $\nu{=}1/3$ state~\cite{Saminadayar97, Dolev08}. Moreover, recent experiments~\cite{Bartolomei20, Nakamura20} have confirmed that the quasihole and quasiparticle excitations of the Laughlin state are Abelian anyons with a statistical phase $\theta{=}2\pi/3$.

\subsection{Magnetoroton excitation}
\label{sec:magnetoroton}

Along with charged excitations, any FQH system, including the Laughlin states, also hosts ``neutral" excitations which can be viewed as arising from a combination of an equal number of quasiparticles and quasiholes. Girvin, MacDonald, and Platzman (GMP)~\cite{Girvin85} proposed a general ansatz for the lowest-lying neutral excitation as a density wave riding on the ground state. The GMP wave function is written as 
$    \Psi^{\rm GMP}_{\nu, \mathbf{k}} {=} \bar{\rho}_{\mathbf{k}}\Psi_{\nu}$,
where $\bar{\rho}_{\mathbf{k}}$ is the LLL projected density operator and $\Psi_{\nu}$ is the ground state wave function at filling $\nu$. The GMP excitation is termed a \emph{magnetoroton} since its dispersion has a characteristic minimum at a finite wavenumber reminiscent of the roton minimum in the excitation spectrum of liquid Helium. The GMP ansatz gives a reasonable account of the exact low-lying neutral excitation at wavenumbers near and below the roton minimum, $k\ell{\lesssim}1$~\cite{He94}. Resonant inelastic light-scattering experiments have mapped out the magnetoroton branch of excitations at $\nu{=}1/3$~\cite{Pinczuk93}.

\subsection{Edge excitations}
\label{sec:edge}

As we sketched in Fig.~\ref{fig:incompressible}(a), due to the insulating bulk, the conduction in FQH states occurs in the vicinity of the system's boundary via chiral edge channels. In the FQH effect, electron correlations are strong and one must model the edge channels using the Luttinger liquid description~\cite{Wen90}. Theoretical description of FQH edge states is provided by bosonization, which conveniently expresses the electronic degrees of freedom in terms of bosons. Due to chirality, backscattering in wide samples is minimized and the edge channels are found to propagate ballistically across long distances, with the dispersion $\epsilon_k {=} v k$, where $v$ is the edge speed and $k$ is the momentum along the edge.

In real samples, however, the edge may undergo a phenomenon of ``reconstruction"~\cite{Chklovskii92, Chamon94} where instead of a single channel, the edge consists of compressible and incompressible stripes that form multiple parallel conducting channels, possibly counterpropagating. Unlike the insulating bulk, the physics of the edge is thus far more prone to non-universal effects due to the long-range tail of the Coulomb interaction and details of the confining potential near the boundary of the system.

The edge spectral function can be measured by tunneling an electron laterally into the edge. Wen's theory~\cite{Wen90} predicts a nonlinear current-voltage, $I{\propto}V^{m}$, relation at zero temperature, with the edge exponent taking the quantized value $m{=}2p{+}1$ for the $\nu{=}1/(2p{+}1)$ Laughlin states. Thus, the quantized edge exponent distinguishes a FQH edge from an ordinary Luttinger liquid, whose edge exponent varies continuously with the strength of the interaction. The edge exponent was measured in experiments by Chang \emph{et al.}~\cite{Chang96} who found its value to be reasonably close to 3 in the $\nu{=}1/3$ state. However, various non-universal effects have also been observed in these experiments~\cite{Chang03}.  This makes the study of FQH edges challenging even in the simplest fractions such as $\nu{=}1/3$.

\subsection{Parent Hamiltonian and Haldane pseudopotentials}
\label{sec:parentHam}

An important step in validation of Laughlin's theory was Haldane's identification of a parent Hamiltonian for the state in Eq.~(\ref{eq: Laughlin_gs})~\cite{Haldane83}. Any homogeneous and isotropic pairwise interaction potential $V(|z_i{-}z_j|)$, with the particles restricted to a single LL, can be written as~\cite{HaldanePrange}
\begin{eqnarray}
H_{\mbox{\scriptsize  int}} = \sum_{i < j} V(|z_i -  z_j|) =\sum_m \sum_{i< j}  V_m \hat P_{ij}^m, \label{eq:HaldanePP}
\end{eqnarray}
where $\hat P_{ij}^m$ projects onto a state in which particles $i$ and $j$ have relative angular momentum $m$. In the LLL, $\hat P_{ij}^m$ projects on the relative state $\psi_\mathrm{rel} {\sim} (z_i {-} z_j)^m$ of any pair of particles. The numbers $V_m$ are called the Haldane pseudopotentials and they represent an energy penalty for two particles to form a state with relative angular momentum $m$. For spinless electrons, only odd values of $m$ can appear as $\psi_\mathrm{rel}$ must be fully antisymmetric in electrons' coordinates. 

The parent Hamiltonian of the $\nu{=}1/3$ Laughlin state is obtained by setting $V_1{>}0$ and all other $V_{m\geq 3}{=}0$. This interaction assigns positive energy to any wave function that vanishes as $\psi_\mathrm{rel}  {\sim} (z_i {-} z_j)$, but it allows pairs of particles to form the state $\psi_\mathrm{rel}  {\sim} (z_i {-} z_j)^3$ for free. The latter is just the $p{=}1$ Laughlin state in Eq.~\eqref{eq: Laughlin_gs}, which emerges as the unique and densest state with exactly zero energy for the $V_1$ pseudopotential interaction. The quasihole excitations of the Laughlin state, Eq.~(\ref{eq: Laughlin_qh}), similarly appear as exact zero-energy states of the same Hamiltonian. Numerics on finite systems show that any other state in the spectrum is separated by a gap on the order $\Delta \propto V_1$ but the persistence of this spectral gap in the thermodynamic limit has not been rigorously proven.

Haldane pseudopotentials greatly facilitate numerical modelling of FQH systems. For example, open-source libraries for exact diagonalization of FQH systems such as DiagHam~\cite{diagham} take full advantage of the pseudopotential formalism. On the analytical side, an arbitrary combination of pseudopotentials can be conveniently converted to a real space potential~\cite{Trugman85}, which enables systematic studies of parent Hamiltonian of more complex FQH states~\cite{Simon07}. 

\section{Hierarchy and composite fermions}
\label{sec: HH_CF}

Following the discovery of the $\nu{=}1/3$ state,  a zoo of more than a hundred FQH plateaux have been observed, many of which do not fit Laughlin's theory (see Fig.~\ref{fig:fqhetraces}). These fractions predominantly belong to the sequence $\nu{=} n/(2pn{\pm}1)$ [and their hole-conjugates at $1{-}n/(2pn{\pm}1)$], where $n$ and $p$ are positive integers (see Fig.~\ref{fig:fqhetraces}). Consequently, new theoretical approaches had to be developed to describe such states. In this chapter we focus on two leading paradigms for understanding the various fractions observed the lowest LL: the Haldane-Halperin hierarchy and Jain's composite fermion theory.

\subsection{Haldane-Halperin hierarchy}
\label{sec: HH_hierarchy}

\begin{figure}[tbh]
	\centering
	\includegraphics[width=0.9\linewidth]{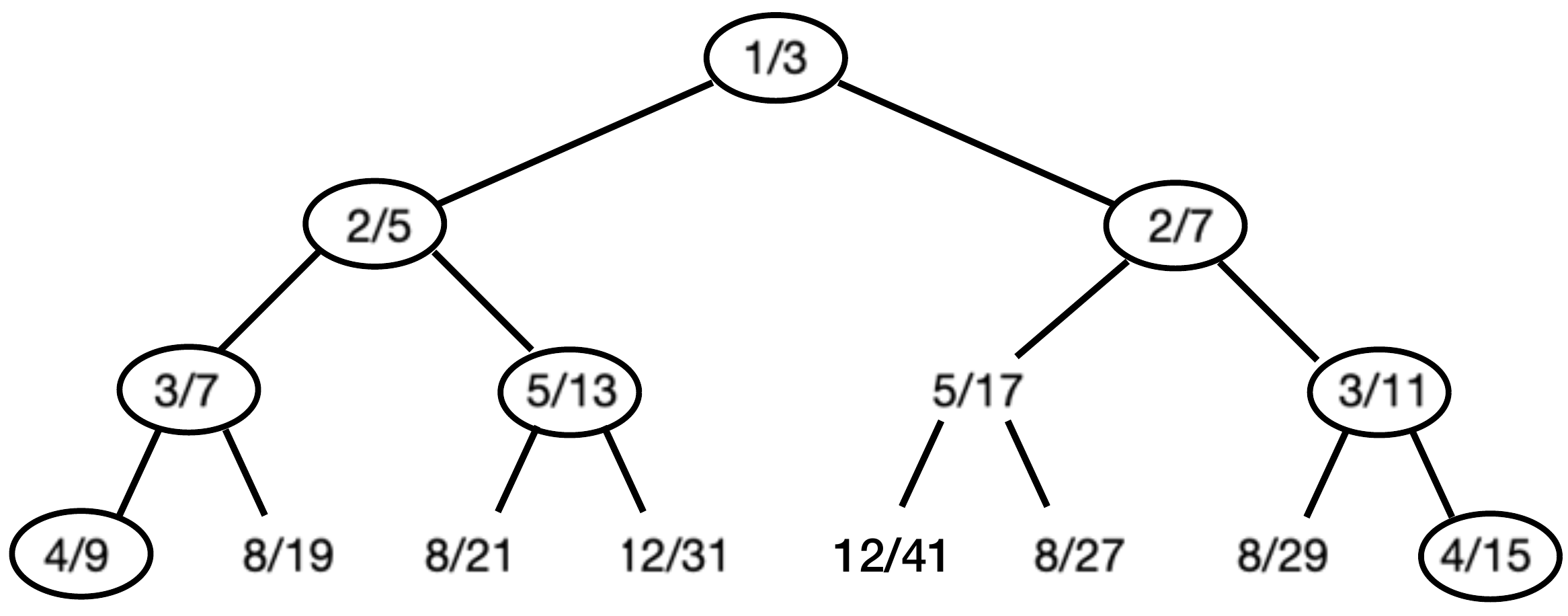}
	\caption{Daughter fractions obtained from the Haldane-Halperin hierarchy originating from the parent Laughlin state at $\nu{=}1/3$. Experimentally observed fractions in the lowest Landau level are circumscribed. }
	\label{fig: Haldane_Halperin_hierarchy}
\end{figure}

Haldane~\cite{Haldane83} and Halperin~\cite{Halperin84} proposed a hierarchical construction to obtain FQH states at arbitrary odd-denominator fractions. In the hierarchical construction, one starts with a nearby parent Laughlin state and views the filling factor difference from it as stemming from the presence of additional Laughlin quasiparticles/quasiholes. Assuming that interaction between the additional Laughlin quasiparticles/quasiholes is sufficiently strongly repulsive at short distances and weak in comparison to the parent Laughlin state, one condenses these excitations into Laughlin-like ``daughter" FQH states. Iterating such a construction, one can produce candidate FQH states at all odd-denominator fractions -- see Fig.~\ref{fig: Haldane_Halperin_hierarchy} for the fractions originating from the $\nu{=}1/3$ Laughlin state in the first three generations of a hierarchy.

A trial wave function for the first daughter state at filling factor $\nu{=}1/(2p{+}1-q^{-1})$, where $q$ is a positive even integer, with $N$ electrons and $N/q$ Laughlin quasiparticles is constructed as follows:
\begin{equation}
    \int \left[\prod_{k=1}^{N/q}d^{2}\eta_{k}\right] [\Phi_{1}^{*}(\{ \bar{\eta}_k \})]^{q}~\Psi^{\rm Laughlin, qps}_{1/(2p+1)}(\{z\};\{\eta_{k}\}), 
\end{equation}
where $\Phi_{1}$ is the usual Laughlin-Jastrow factor and the term  $\Psi^{\rm Laughlin, qps}_{1/(2p{+}1)}$ is the Laughlin quasiparticle state of Eq.~\eqref{eq: Laughlin_qp} generalized to multiple quasiparticles located at positions $\eta_{k}$. Analogous wave functions for the first daughter state at $\nu{=}1/(2p{+}1+q^{-1})$ can be constructed from  Laughlin's quasihole wave function of Eq.~\eqref{eq: Laughlin_qh}. The same idea can be used to construct wave functions for daughter states further down in the hierarchy. 

For small systems at $\nu{=}2/5$ and $3/7$, the hierarchical wave functions have been constructed numerically and these have high overlaps with the exact LLL Coulomb ground states~\cite{Greiter94}. Studies of larger systems and other filling factors are required to substantively test these wave functions. The order of fractions and their stability predicted by the hierarchy theory is not entirely consistent with experiments. For example, the 3/7 and 5/13 states appear at the same level in the hierarchy (see Fig.~\ref{fig: Haldane_Halperin_hierarchy}) but the FQH effect at $3/7$ is much stronger than that at $5/13$ as is evidenced from their gaps [see Fig.~\ref{fig:fqhetraces}(a)]. Nonetheless, universal aspects, such as the charge and braiding statistics of the quasiparticles, predicted by the Haldane-Halperin construction are consistent with experiments. 

\subsection{Composite fermions}
\label{sec:CF_theory}

Jain's theory~\cite{Jain89} provides a conceptually different microscopic explanation of the $\nu{=}n/(2pn{\pm}1)$ FQH states as IQH states of certain emergent topological particles called \emph{composite fermions} (CFs). A composite fermion is a bound state of an electron and an even number ($2p$) of vortices. Due to the vortex-binding, the CFs see a reduced magnetic field $B^{*}{=}B{-}2p\rho \phi_{0}$ compared to the external magnetic field $B$ seen by the electrons [superscript $*$ is used to denote CF quantities]. The ${+}$ (parallel) and ${-}$ (anti-parallel) signs in $\nu{=}n/(2pn{\pm}1)$ indicate the direction of the residual magnetic field seen by the CFs with respect to the external magnetic field. In the zeroth-order approximation, the CFs are assumed to be non-interacting, and just like free electrons in a magnetic field, CFs form their own Landau-like levels, called \emph{Lambda levels} ($\Lambda$Ls). When an integer number $\nu^{*}{=}n$ of these $\Lambda$Ls are fully filled, we get an FQH effect of electrons at $\nu{=}n/(2pn{\pm}1)$. The existence of $\Lambda$Ls has been established in experiment~\cite{Kang93}. Recently, in an ultra-high mobility sample, FQH states all the way up to $n{=}16$ in the $n/(2n{\pm}1)$ Jain sequence have been observed~\cite{Chung21}.

\begin{figure}[tbh]
	\centering
	\includegraphics[width=\linewidth]{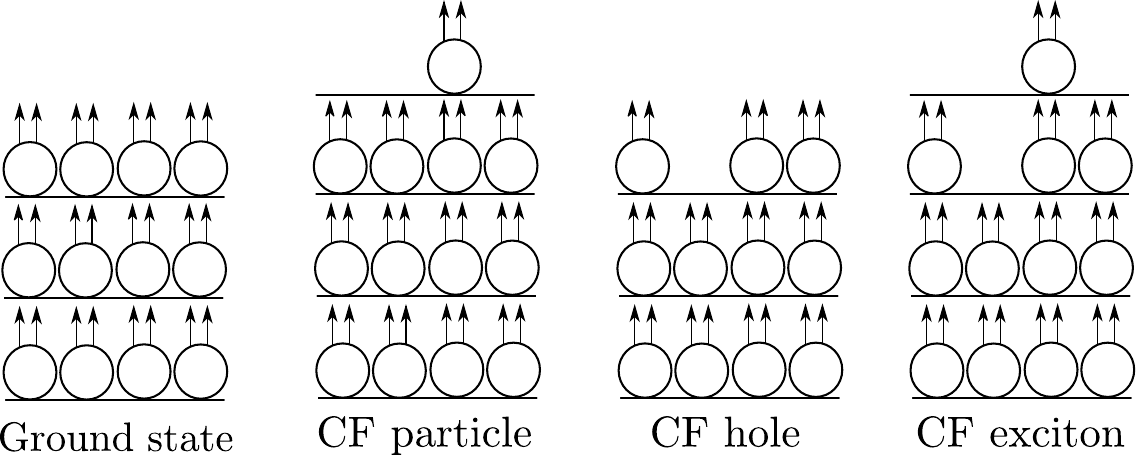}
	\caption{Schematic representation of CF ground state, elementary charged and neutral excitations at $\nu{=}3/7$. }
	\label{fig: CF_all}
\end{figure}

The mapping to an IQH effect leads to the Jain wave function for the FQH ground state at $\nu{=}n/(2pn{\pm}1)$~\cite{Jain89}
\begin{equation}
    \Psi^{\rm Jain}_{n/(2pn\pm 1)} = \mathcal{P}_{\rm LLL}\Phi^{2p}_{1}\Phi_{\pm n}.
    \label{eq: Jain_wf_gs}
\end{equation}
Here $\Phi_{n}$ is the Slater determinant state of $n$ filled LLs of electrons, while the Jastrow factor $\Phi^{2p}_{1}$ attaches $2p$ vortices to each of the electrons to convert them into CFs, see  Fig.~\ref{fig: CF_all} for a schematic illustration. For $n{=}1$ (with the ${+}$ sign) the Jain wave function is identical to the Laughlin wave function of Eq.~\eqref{eq: Laughlin_gs}. For $n{>}1$, the Jain wave function needs to be explicitly projected to the LLL, as denoted by $\mathcal{P}_{\rm LLL}$. 

Beyond ground state properties, the IQH mapping readily allows the construction of wave functions for the excitations as well, see Fig.~\ref{fig: CF_all}. The positively charged quasihole called a CF hole is obtained by removing a CF from the topmost filled $\Lambda$L. The negatively charged quasiparticle, referred to as a CF particle, is obtained by adding a CF in the topmost empty $\Lambda$L. The lowest-lying neutral excitation, called the CF-exciton, is obtained by combining a single CF particle and a CF hole. The excitations obtained using CF theory generally exceed the accuracy of other approaches, e.g., the Laughlin and GMP ansatz discussed in Sec.~\ref{sec:Laughlin_fractions}, in describing the exact LLL Coulomb excited states.  

We note here that although the Haldane-Halperin hierarchical states are topologically equivalent to CF states~\cite{Read90, Wen95}, the two theories propose different microscopic mechanisms for the origin of FQH effect~\cite{Jain14}. Unlike the hierarchy wave functions, the CF wave functions in Eq.~(\ref{eq: Jain_wf_gs}) have proven amenable to large-scale numerical simulations~\cite{Jain98}. These studies have demonstrated that the CF theory provides a  highly-accurate description of the majority of FQH states observed in the LLL. For example, neutral excitation spectrum observed using surface acoustic waves~\cite{Kukushkin09} is in good agreement with CF theory~\cite{Scarola00}. Nevertheless, some quantitative discrepancy between the experimentally measured~\cite{Boebinger85,Willett88,Du93,Pan20,Rosales21} and theoretically predicted~\cite{Zhang86, Ortalano97, Morf02} charge gaps persists to this day. This likely arises from neglecting the effects of the disorder, interface roughening, and screening by gates in theoretical studies. Furthermore, theoretical calculations use simplified models to incorporate the effects of LL mixing and finite-width. More realistic calculations are needed to quantitatively explain the gaps obtained in experimental studies.
Finally, certain delicate states, like those observed at $\nu{=}4/11$ and $5/13$~\cite{Samkharadze15b, Pan15}, Fig.~\ref{fig:fqhetraces}(a), and the states seen in the SLL, Fig.~\ref{fig:fqhetraces}(b), lie beyond the paradigm of \emph{weakly-interacting} CFs presented in this section. To understand such states, one needs other approaches, as discussed in subsequent sections below. 

\section{Half-filled Landau level}\label{sec:halffilling}

The simplest fraction -- half-filling of a LL -- has remained in the focus of research since the early days of  the FQH effect. The  conundrum is that no quantized FQH plateau is observed when it is the LLL that is half-filled, i.e., at $\nu{=}1/2$ -- see Fig.~\ref{fig:fqhetraces}(a). Yet, when the \emph{second} LL is half-filled, resulting in a total filling factor $\nu{=}5/2$, a robust FQH plateau is seen instead [Fig.~\ref{fig:fqhetraces}(b)]. In this section, we discuss a novel kind of gapless Fermi-liquid like state that was introduced describe the $\nu{=}1/2$, and its pairing instability that is believed to  give rise to the exotic gapped state at $\nu{=}5/2$.

\subsection{Gapless composite fermion Fermi liquid in the half-filled Landau level}
\label{sec: CFL}

The  $\nu{=}1/2$ state finds a natural interpretation in the CF theory since at half-filling CFs see no residual magnetic field. Therefore, like electrons at zero field, CFs can form a Fermi liquid state, called the composite Fermi liquid (CFL)~\cite{Kalmeyer92, Halperin93}, which is gapless and thus does not lead to a FQH effect. The CFL is a markedly different state from the ordinary Fermi sea: in the LLL, there is no kinetic energy and interactions alone conspire to create an emergent Fermi-liquid. The wave function for the CFL at even-denominator fillings $\nu{=}1/(2p)$ is obtained as the $n{\rightarrow}\infty$ limit of the wave function in Eq.~\eqref{eq: Jain_wf_gs}:
\begin{equation}
    \Psi^{\rm CFL}_{1/(2p)} = \mathcal{P}_{\rm LLL} \Phi^{\rm FL} \Phi^{2p}_{1},~~\Phi^{\rm FL} \equiv {\rm det}\left[e^{i\vec{k}_{l}.\vec{r}_{m}}\right].
    \label{eq: CFL_wf}
\end{equation}
Halperin, Lee, and Read (HLR)~\cite{Halperin93} evaluated many properties of the CFL using a Chern–Simons field theory in which flux quanta are bound to electrons to turn them into CFs. This description is in qualitative agreement with numerous experiments that have verified the presence of the CFL~\cite{Kang93, Willett93, Goldman94}. In Sec.~\ref{sec:field} we shall discuss one particular experiment, namely the measurement of the CF Fermi wave vector using geometric resonance~\cite{Willett99, Kamburov14b, Hossain20}, which has renewed the interest  in the CFL state.

\subsection{$\nu{=}5/2$ state}
\label{sec: MR_Pf_5_2_FQHE}

The observation of a FQH state at the even-denominator filling  $\nu{=}5/2$~\cite{Willett87} indicated that not all FQH states can be described using the ideas discussed in Sec.~\ref{sec: HH_CF}. Moore and Read~\cite{Moore91} identified conformal field theory (CFT) as an innovative tool for generating trial FQH wave functions (see Ref.~\cite{HanssonRMP} for an in-depth overview of this technique). Using a particular type of CFT, Moore and Read arrived at the following wave function for a state at $\nu{=}1/2$:
\begin{eqnarray}\label{eq:MR}
	\Psi_{1/2}^{\mathrm{MR-Pf}}=\operatorname{Pf}\left(\frac{1}{z_{i}-z_{j}}\right) \prod_{1\leq i<j \leq N}\left(z_{i}-z_{j}\right)^2,
\end{eqnarray}
where the number of particles $N$ is assumed to be even. The key component of this wave function is the Pfaffian  -- the antisymmetrized sum over pairs:
\begin{eqnarray}
	\operatorname{Pf}\left(\frac{1}{z_{i}-z_{j}}\right)=\mathbb{A}\left(\frac{1}{z_{1}-z_{2}} \frac{1}{z_{3}-z_{4}} \cdots\right), 
\end{eqnarray}
where $\mathbb{A}$ represents the anti-symmetrization operator. The Pfaffian factor makes the overall wave function antisymmetric and therefore a valid trial state for electrons. As first suggested by Greiter, Wen and Wilczek~\cite{Greiter92}, this wavefunction can provide a description of the experimental $\nu{=}5/2$ resistance plateau, provided one assumes that the two lowest LLs of $\uparrow$ and $\downarrow$ spins are completely filled with electrons. Thus, Eq.~\eqref{eq:MR} describes the correlations between fully spin-polarized electrons in the valence (half-filled) SLL. 

The Pffafian factor introduces weak pairing between the CFs~\cite{Read00}. Indeed, in the antisymmetrized product over pairs, we recognize the real-space form of the wave function describing a 2D Bardeen-Cooper-Schrieffer (BCS) superconductor. Thus, we can view Eq.~\eqref{eq:MR} as the FQH analog of a $p_{x}{+}i p_{y}$ superconductor. The physical origin of pairing is due to the effective Coulomb interaction being less repulsive at short distances in the SLL compared to LLL. Thus, forming CFs overscreens the Coulomb repulsion resulting in a weak residual attractive interaction between them which leads to a pairing instability of their CFL. One consequence of the electron pairing is an unconventional collective excitation on top of the Moore-Read ground state that was dubbed the ``neutral fermion" mode~\cite{Bonderson2011, Moller2011}.

The interest in the Moore-Read state was initially fuelled by numerical simulations. Exact diagonalizations performed by Morf~\cite{Morf98} showed that the spin-polarized state at $\nu{=}5/2$ had lower energy than any partially polarized or spin-singlet state, even in the absence of any Zeeman splitting. Furthermore, it was shown that the state was poised towards a compressible phase~\cite{Rezayi00} and can be easily pushed across the phase boundary by slightly varying the Hamiltonian in a way that mimics the effect of an in-plane magnetic field, accounting for experimental observations~\cite{Eisenstein90,Lilly99, Pan1999}. The correspondence between numerics and experiment has been made more quantitative by comparisons of the energy gap obtained from exact diagonalization~\cite{Morf02} and DMRG simulations~\cite{Feiguin2008} with the one measured in experiments~\cite{Kumar10}.  

\subsection{Anti-Pfaffian}\label{sec:apf}

While PH symmetry is exact for the FQH problem restricted to a single LL, the Moore-Read state in Eq.~\eqref{eq:MR} is not invariant under PH transformation. The state obtained by PH conjugation applied to Eq.~\eqref{eq:MR} is a topologically-distinct phase of matter that was named the ``anti-Pfaffian" (aPf)~\cite{Levin07, Lee07}. The two states have different thermal Hall responses and different edge excitations, which immediately leads to the question: which state describes the $\nu{=}5/2$ ground state in the thermodynamic limit? 

Distinguishing between the aPf and Moore-Read states has proven a considerable challenge. The key to discriminating between the two  states is LL mixing, which explicitly breaks PH symmetry and lowers the energy of one of these states with respect to the other. Unfortunately, accurate modeling of LL mixing effects comes with many challenges, in particular its perturbative treatment~\cite{Bishara09, Sodemann13, Simon13, Pakrouski15} cannot be rigorously justified. Nevertheless, as recently shown by Rezayi~\cite{Rezayi2017}, different theoretical approaches converge to the conclusion that LL mixing favors the aPf state. 

On the other hand, experimental studies have detected signatures of $e/4$ quasiparticles at $\nu{=}5/2$ using the interference of edge channels~\cite{Willett09}. Shot noise measurements~\cite{Dolev10} and charge-sensing in the bulk using single-electron transistor~\cite{Venkatachalam11} have also reported evidence of $e/4$ quasiparticles. These results are consistent with both Moore-Read and aPf states. The dependence on voltage and temperature of the tunneling current was measured at a point contact in the $\nu{=}5/2$ state and found to be consistent with the aPf state~\cite{Radu08}. Furthermore, upstream neutral modes have been detected at $\nu{=}5/2$~\cite{Bid10, Dolev11} which is consistent only with the aPf state.

\section{Non-Abelian states}
\label{sec:nonabelian}

Contrasting the case of SLL with that of the LLL in Fig.~\ref{fig:fqhetraces}(a)-(b), one notices a few striking differences. One of the most prominent incompressible states in the SLL has an even denominator, $\nu{=}2{+}1/2$~\cite{Willett87} and its strength is roughly comparable with $\nu{=}2{+}1/3$, which is expected to be analogous to the $\nu{=}1/3$ Laughlin state in the LLL~\cite{Balram13b}. Moreover, FQH effect is also observed at $\nu=2{+}3/8$~\cite{Xia04, Kumar10}. Even the odd-denominator FQH states in the SLL appear out of order when compared with the LLL sequence: FQH effect has been observed at $1/3$ and $2/5$ in the LLL, but no FQH effect has been confirmed at $3/7,~4/9$ and $5/11$ but evidence for a plateau at $2{+}6/13$ exists~\cite{Kumar10}. To explain these unconventional fractions, different theoretical approaches have been put forward. A common theme in these approaches are quasiparticles with non-Abelian braiding statistics, qualitatively different from the Abelian anyons we encountered in Sec.~\ref{sec: incompressible_fluids}.

\subsection{Non-Abelian anyons and topological quantum computation}\label{sec:tqc}

Beyond the paired ground state wave function, the quasihole excitations of the Moore-Read state in Eq.~\eqref{eq:MR} are particularly interesting: they must be created in pairs and they have non-Abelian braiding statistics. Imagine that $2n$ quasiholes have been added to the Moore-Read state by increasing the magnetic flux by $n$ flux quanta. In non-Abelian FQH states, fixing the configuration of the quasiholes does not result in a unique quantum state -- instead, there is a degenerate manifold of states. For example, in the Moore-Read state the $2n$-quasihole manifold is $2^{n{-}1}$-fold degenerate and is spanned by some orthogonal basis $\{\psi_1,\psi_2,{\cdots}, \psi_{2^{n{-}1}}\}$. The braiding of two quasiholes, depicted in Fig.~\ref{fig:incompressible}(c), acts as a unitary transformation on this subspace, $	\psi_\alpha {=} \mathcal{U}_{\alpha\beta}\psi_\beta$~\cite{Nayak96}. As matrices $\mathcal{U}$ do not generally commute, the quasiholes in the Moore-Read state obey `non-Abelian' braiding statistics. 

The degenerate states $\psi_\alpha$ can be interpreted as basis states of a logical qubit, and the quasihole braiding operations $\mathcal{U}$ can be mapped to the action of quantum gates~\cite{Freedman03}. 
These operations are topologically protected at sufficiently low temperatures $k_B T {\lesssim} \Delta_\mathrm{MR}$, where $\Delta_\mathrm{MR}$ is the excitation gap of the Moore-Read state. This is because the qubit states are locally indistinguishable and any decoherence induced by local perturbations will be strongly suppressed. This idea of employing a topologically-protected manifold of non-Abelian FQH quasiholes to encode and manipulate a quantum bit is known as \emph{topological quantum computation}~\cite{Nayak08}. 

While the prospect of avoiding the need for error correction in a quantum computer is of immense practical benefit, the particular system of Moore-Read anyons discussed above falls short of this goal as it is unable to perform so-called universal quantum computation.. The simplest FQH system capable of universal topological quantum computation is discussed in Sec.~\ref{sec:125}. At this stage, however,  topological quantum computation with FQH states remains a theoretical concept, although recent progress in FQH interferometry~\cite{Willett2021, Nakamura20} paves the way to a potential demonstration of an operating FQH qubit. We note that special types of one-dimensional nanowires can host boundary excitations with properties reminiscent of Moore-Read anyons~\cite{Kitaev01}, offering a potentially more practical realization of topological quantum computation. The efforts towards topological quantum computation will rely critically on better understanding and better control of FQH excitations, which may be aided by STM techniques~\cite{Papic2018}.

\subsection{$\nu{=}12/5$ state}\label{sec:125}

Universal quantum computation can be performed with more complex types of anyons that emerge in the Read-Rezayi (RR) sequence of states~\cite{Read99}. These non-Abelian states generalize the notion of pairing between pairs of particles in the Moore-Read state to a clustering of larger groups (three or more particles). The RR states occur at filling factors $\nu{=}k/(k{+}2)$, with $k{=}1,2$ being the Laughlin and Moore-Read states, respectively. The $k{=}3$ RR state is a qualitatively new non-Abelian state hosting the so-called Fibonacci anyons. This state is often referred to as $\mathbb{Z}_3$ RR state because its wavefunction was originally constructed using a  $\mathbb{Z}_3$ parafermion conformal field theory~\cite{Read99}.

Experimental observation of the $\nu{=}12/5$ FQH plateau by Xia \emph{et al.}~\cite{Xia04} and subsequently Kumar \emph{et al.}~\cite{Kumar10} has generated much excitement due to the possibility that it may be described by the PH conjugate of the $\mathbb{Z}_3$ RR state. Numerical work~\cite{Read99} indicated that the Coulomb ground state at $\nu{=}12/5$ is close to a phase transition between the Abelian CF state and the PH conjugate of $\mathbb{Z}_3$  RR state. More recent studies using DMRG~\cite{Mong15} have found that the ground state at $\nu{=}12/5$ is fully spin-polarized and in the vicinity to a charge-density-wave ordered phase. The latter was found to be much more sensitive to LL mixing at $\nu{=}13/5$ than at $\nu{=}12/5$~\cite{Pakrouski16}, accounting for the absence of a quantized plateau at $\nu{=}13/5$. Due to a much smaller gap in the $\nu{=}12/5$ state and the scarcity of its experimental realizations, the definitive determination of its nature remains much more challenging compared to $\nu{=}5/2$.

\subsection{Other non-Abelian states}

Previously mentioned RR states are a small subset of a family of states associated with the so-called Jack polynomials~\cite{FeiginJimboMiwa}. As first realized by Bernevig and Haldane~\cite{Bernevig08}, the wave function of many FQH states at filling factors $\nu{=}k/(k{+}r)$ can be identified  with a single Jack polynomial, parametrized by integers $k$, $r$ and a so-called ``root" state. In the case $r{=}2$ the Jacks coincide with the RR states discussed in Sec.~\ref{sec:125}. An interesting case with $k{=}2$ and $r{=}3$ corresponds to the so-called ``Gaffnian" state~\cite{Simon07b}, which has inspired much theoretical work as a potential example of a gapless non-Abelian state~\cite{Read09a}.
The recursive properties of the Jack polynomials impose a great deal of analytical structure on FQH wave functions, a fact that has been fruitfully employed in numerical computations~\cite{Bernevig09}.

Bonderson and Slingerland~\cite{Bonderson08} constructed trial wave functions for many of the filling factors where FQH states have been observed in the SLL. The Bonderson-Slingerland approach obtains FQH wave functions as a product of the bosonic version of the Moore-Read state and a composite fermion state. In particular, this approach yields a state that is competitive with the (PH conjugate of) $\mathbb{Z}_3$ RR state at $\nu{=}12/5$~\cite{Bonderson12}. On the other hand, the observed filling factor $2{+}6/13$~\cite{Kumar10} is not a part of this sequence. Other constructions, based on condensing non-Abelian quasiparticles of the Moore-Read state~\cite{Hermanns10} or by taking products of the bosonic RR states with the Laughlin state~\cite{Jolicoeur07}, have also been proposed. The hierarchical construction of Levin and Halperin~\cite{Levin09a} may account for the FQH effect observed at filling $\nu{=}2{+}6/13$. Somewhat surprisingly, the Levin-Halperin states turn out to be Abelian, even though they derive from a non-Abelian Moore-Read parent state. Most of these states, however, result in complicated wave functions that have not been systematically tested in numerical simulations.

\section{Multicomponent fractional quantum Hall effect}\label{sec:multi}

\begin{figure*}
    \includegraphics[width=\linewidth]{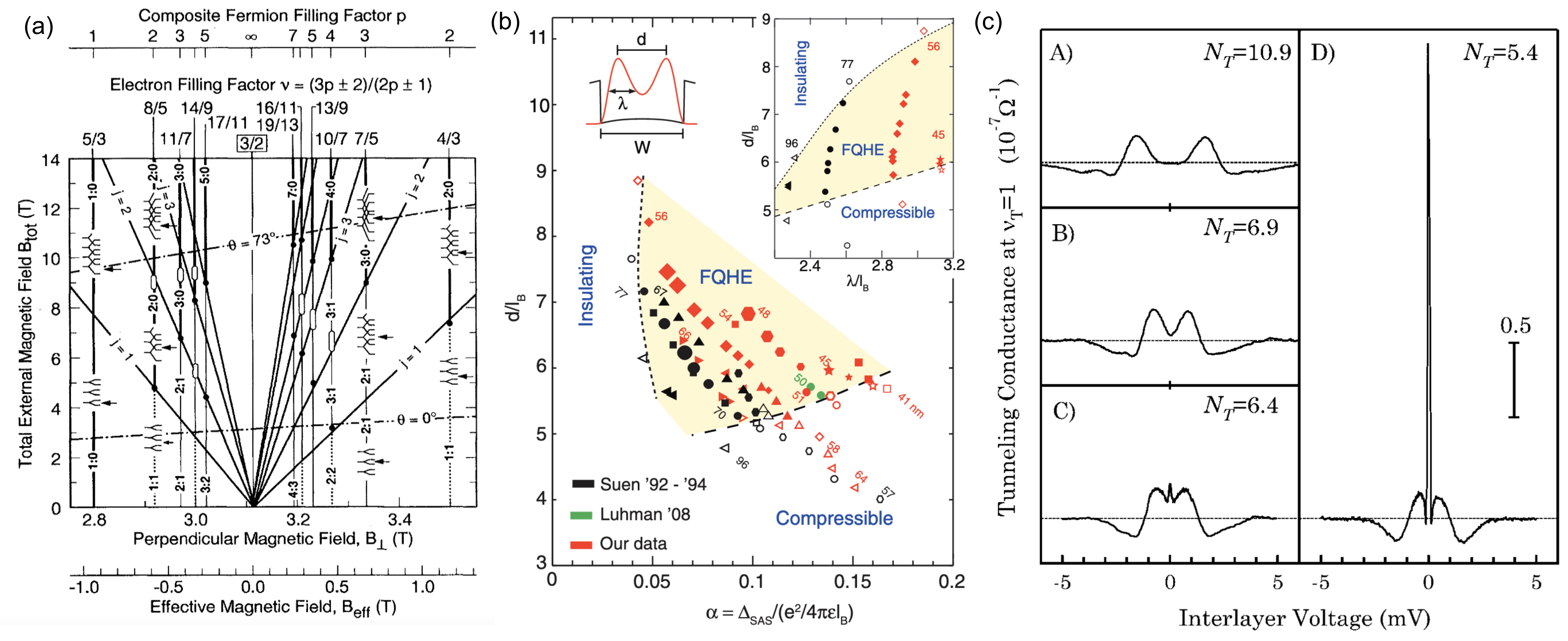}
    \caption{Multicomponent FQH effects. 
    (a) Transitions between CF states with different spin polarization (solid dots) as in-plane magnetic field is varied. The lines are fits from the free CF theory, with the $g^{*}$ factor and the effective mass $m^{*}$ treated as empirical parameters. $B_{\text {eff }}$ is the effective magnetic field. Data from Ref.~\cite{Du95}.
    (b) Phase diagram of GaAs wide quantum wells at $\nu_T{=}1/2$ as a function of  $d / \ell$ and $\alpha{=}\Delta_{\mathrm{SAS}} /\left(e^{2} / 4 \pi \epsilon \ell\right)$. Region corresponding to an incompressible FQH state is shaded in yellow. The interlayer distance $d$ and layer width $\lambda$ are estimated numerically from the self-consistent charge distribution. Figure reproduced from Ref.~\cite{Shabani13}.
    (c) Exciton condensation in $\nu_{T}{=}1$ bilayers. 
    Tunneling conductance $dI/dV$ vs interlayer voltage $V$ for a range of total densities $N_{T}$, corresponding to $1.6 \lesssim d/\ell {\lesssim} 2.3$. Trace $A$, at the highest density, shows a deep suppression of the tunneling near zero bias. By trace $D$, the lowest density of the four shown, this suppression has been replaced by a tall peak, signalling exciton condensation. The vertical scale is the same for all traces. Figure reproduced from Ref.~\cite{Spielman00}. 
    }
    \label{fig:multi}
\end{figure*}

Up to this point, our discussion was restricted to fully spin-polarized electrons, as is appropriate for the setting when the external magnetic field is large. It turns out that the Zeeman splitting $E_{Z}$ (see Sec.~\ref{sec:intro}) is strongly suppressed in GaAs: the ratio of $E_Z$ to the Coulomb energy scale for a typical magnetic field $B{=}9\mathrm{T}$ is $\xi{=}E_{Z}/(e^{2}/(4\pi\epsilon\ell)){\approx} 0.02$. Experimentally, it is possible to tune $\xi$ by tilting the sample or introducing a parallel magnetic field, suppressing the $g$-factor using hydrostatic pressure~\cite{Leadley97, Nicholas98} or changing the 2DEG's density. Therefore, for certain parameter regimes, it could be favorable for a fraction of the electrons to flip their spin leading to the formation of unpolarized states~\cite{Chakraborty84, Chakraborty00}. We now turn our attention to such ``multicomponent" FQH states where the internal degree of freedom allows the electrons to form new kinds of correlated states without analogs in single-layer FQH systems.

Apart from ordinary spin, the valley degree of freedom could also lead to multi-component FQHE and this plays a particularly important role in materials such as graphene. Here, we just mention that multi-component FQHE has been also observed in multi-valley semiconductor systems such as AlAs~\cite{Bishop07, Padmanabhan09}. Furthermore, it has also been possible to fabricate two single-layer 2DEGs in close proximity to each other, such that the two layers assume the role of $\uparrow$, $\downarrow$ spin states. The advantage of studying such bilayer FQH systems is the enhanced tunability of the interactions: one can vary both the  tunneling $\Delta_{SAS}$ between the two layers, as well as the effective interlayer distance $d/\ell$. Note that, conveniently, the physical distance $d$ can be fixed and one varies $\ell$ by tuning the overall magnetic field (while also adjusting the electron density to remain at the same filling factor). The parameter $d/\ell$ directly modifies the interlayer Coulomb interaction, which has a strong effect on many-electron states. In experiment, two main architectures were pursued: wide single quantum wells in which electrostatic effects produce a ground subband wave function with a pronounced dumbbell shape~\cite{Suen92} and true double quantum wells consisting of two thin GaAs layers embedded in the alloy $\mathrm{Al}_{x} \mathrm{Ga}_{1{-}x} \mathrm{As}$~\cite{Eisenstein92}.

Theoretical understanding of multicomponent FQH states is guided by Halperin's generalization of the Laughlin wave function in Eq.~(\ref{eq: Laughlin_gs}) to FQH systems with a discrete internal quantum number~\cite{Halperin83}. Halperin's states are parametrized by non-negative integers $m$, $m'$ and $n$ and their wave functions are given by
\begin{equation}
	\Psi^{(m,m',n)} = \prod_{i<j}(z_{i}-z_{j})^{m}\prod_{i<j}(w_{i}-w_{j})^{m'}\prod_{i,j}(z_{i}-w_{j})^{n},
	\label{eq:Halperin_mmn}
\end{equation}
where $\{z_{k}\}$ and $\{w_{k}\}$ are the coordinates of the electrons in the ``top" and ``bottom" layer  (or spin state), respectively. The exponents $m$ and  $m'$ must be odd integers due to Pauli exclusion principle, while $n$ does not have such a restriction. Moreover, because  intra-species correlations are typically stronger than inter-ones we must have $m, m' {\geq} n$. The state in Eq.~(\ref{eq:Halperin_mmn}) corresponds to the filling factor $\nu_T = (m{+}m'{-}2 n)/(m m'{-}n^{2})$. Most of the time, one is interested in the balanced case where the two layers or spin species are equivalent, i.e., $m=m'$, which can be physically interpreted as each layer forming a $\nu{=}1/m$ Laughlin state, with interlayer correlations controlled by exponent $n$. 

We note that the Halperin $(m,m,n)$ states can be generalized to bilayer CF states, where the $\nu{=}1/m$ Laughlin states in each layer are replaced by a Jain state~\cite{Scarola01b}. Furthermore, in the case of spin, the CF wave function in Eq.~\eqref{eq: Jain_wf_gs} can be directly generalized to accommodate the spin degree of freedom by filling up $n_{\uparrow}$ LLs with $\uparrow$-spin electrons (and similarly $n_{\downarrow}$ LLs filled with electrons of opposite spin)~\cite{Wu93}. Depending on $n_{\uparrow}$ and $n_{\downarrow}$, states with full or partial spin polarization can be readily constructed. 

\subsection{Spinful systems}
\label{sec:spin}
In single-layer FQH systems, as $\xi$ is varied, transitions between states with different spin polarization can take place and have been extensively studied in experiment, see an illustrative example in Fig.~\ref{fig:multi}(a). For example, at $\nu{=}4/7$,three CF states can be constructed: a) fully polarized with $n_{\uparrow}{=}3$ and $n_{\downarrow}{=}0$, b) partially polarized with $n_{\uparrow}{=}3$ and $n_{\downarrow}{=}1$, and c) spin-singlet with $n_{\uparrow}{=}2{=}n_{\downarrow}$. At large $\xi$, the ground state is fully polarized. As $\xi$ is lowered, the fully polarized state transitions to a partially polarized state and eventually at small $\xi$ goes into a spin-singlet state. At the Laughlin fractions, even at zero Zeeman splitting, the ground state is fully spin polarized due to exchange interactions, an effect that has been dubbed quantum Hall ferromagnetism~\cite{Sondhi93}.

In the presence of spin, PH symmetry relates states at $\nu$ to $2{-}\nu$. Experimentally in GaAs, spin transitions have predominantly been studied at fractions in the range $1{<}\nu{<}2$~\cite{Clark89, Du95}. The reason for this is at larger fillings, the magnetic field is smaller and thus the starting Zeeman energy is lower which allows access to the least polarized states. Then by tilting the sample, Zeeman energy can be increased and states with higher polarizations can be accessed. The spin phase diagram of the Jain fractions has been worked out in detail both theoretically~\cite{Park98, Balram15} and experimentally~\cite{Du95,Kukushkin99, Liu14} and these are in semi-quantitative agreement with each other. 

In the presence of spin, the excitations of fully polarized ground states can also change character. In Secs.~\ref{sec:magnetoroton} and \ref{sec:CF_theory} we discussed excitations which are spin-conserving. Howver, at low values of $\xi$, the lowest-lying excitation could involve a spin-flip. At the Laughlin fractions, this lowest-lying spin-flip mode, termed the spin-wave, has energy identical to the Zeeman splitting in the long-wavelength limit~\cite{Kallin84}. In the fully polarized CF states the spin-wave mode has a roton minimum~\cite{Mandal01} which lies below the ground state at zero-Zeeman splitting. The presence of these low-lying excitations is indicative of the fact that as $\xi$ is lowered the fully polarized ground state could become unstable giving way to another state with lower polarization.

More exotic excitations involving spin-flip such as skyrmions~\cite{Sondhi93, Fertig94}, that carry a topological spin texture, can also be stabilized at low values of $\xi$ in the vicinity of a Laughlin fraction. These skyrmions can be viewed as CF particles or CF holes dressed by spin waves. The exchange interaction between CFs is much weaker than that between electrons, so it is a lot harder to realize and detect fractional skyrmions compared to their integer counterparts. Nevertheless, Zeeman-energy dependence of certain excitations does point to the existence of neutral spin-textures in the vicinity of $\nu{=}1/3$ which have been interpreted as arising from pair(s) of skyrmions and anti-skyrmions~\cite{Groshaus08}. More direct evidence for their existence has been obtained from the measurement of the binding energy of fractional skyrmions in the vicinity of $\nu{=}1/3$ using light-scattering experiments~\cite{Balram15d}.

\subsection{Bilayer systems}\label{sec:bilayer}

The first experimental evidence for unusual states in bilayer systems was the observation of a quantized Hall plateau at $\nu_T {=}1/2$ in a bilayer system with  $d / \ell {\lesssim} 3$, in stark contrast to a single-layer FQH system at the same filling~\cite{Suen92,Eisenstein92}. Numerical calculations~\cite{Chakraborty87,Yoshioka89,He93}, verified the existence of an incompressible state and identified it with the (3,3,1) Halperin state in Eq.~(\ref{eq:Halperin_mmn}). More recently, there has been some renewed interest in the $\nu_T {=}1 / 2$ two component systems~\cite{Shabani13,Shabani09b} due to the possible transition into the Moore-Read state as tunneling is increased~\cite{Ho95, Peterson10,Zhao20}. An example of the phase diagram in a wide quantum well at $\nu_T{=}1/2$ is reproduced in Fig.~\ref{fig:multi}(b).  

Perhaps the most remarkable bilayer quantum Hall phenomenon is the exciton condensation observed at total filling $\nu_T {=}1$~\cite{Spielman00,Tutuc04}. The key to demonstrating the existence of excitons was the measurement of interlayer tunneling conductance at zero bias, which revealed a sharp change in the behavior depending on $d / \ell$.  When $d / \ell{=}$ is sufficiently small, one observes a pronounced Josephson-like peak~\cite{Spielman00}, see Fig.~\ref{fig:multi}(c). The peak reflects the fact that there is no energy cost associated with the transfer of an electron from one layer to the other. Furthermore, one can also probe the system in a so-called ``drag" configuration, where the current is driven through one of the layers and the voltage drop is measured in the other layer. In the exciton phase, the Hall drag resistance becomes accurately quantized~\cite{Kellogg02,Tutuc2009}. 

Theoretical explanation of these experimental findings is provided by the Halperin (1,1,1) state of Eq.~\eqref{eq:Halperin_mmn}. In this state, each electron is a coherent linear superposition of the two layers~\cite{Fertig1989} where the relative phase $\phi$ is arbitrary and the same for all electrons. Such a state is ferromagnetically ordered in the $x{-}y$ plane, inclined by the angle $\phi$ relative to the $x$-axis and spontaneously chosen by the system. As a consequence of spontaneous U(1) symmetry breaking, the system acquires a gapless Goldstone mode -- a  linearly dispersing pseudospin wave~\cite{Moon95}, which was detected in Ref.~\cite{Spielman2001}. 

While the properties of the $\nu_T{=}1$ bilayer have been well-understood in the limits of small and large $d/\ell$, the nature of the transition at some critical $d_c/\ell$ and a possibility of intermediate phases remain the subject of on-going work~\cite{ Moller2008,  Milovanovic2015, Zhu2017, Lian2018, Wagner2021}.

\section{Recent developments}

In recent years the interest in FQH phases has expanded  beyond  topology into broader investigations of geometric effects and quantum entanglement. Moreover, there has been a resurgence of interest in the parton description of FQH states and effective field theories that incorporate geometric effects and particle-hole symmetry. This section is an overview of these recent developments in the field.

\subsection{Parton theory}\label{sec:parton}

Parton theory was initially introduced as a generalization of CF theory to describe a wider class of states~\cite{Jain89b}. In the parton theory, one imagines splitting the electron into $k$ (an odd integer) fictitious sub-particles called \emph{partons}. 
To obtain an incompressible state, each of the partons is placed in an IQH state (or in general, any known incompressible state), resulting in the wave function:
\begin{equation}
\Psi^{n_{1}n_{2}\cdots n_{k}}_{\nu} = \mathcal{P}_{\rm LLL}\prod_{\beta=1}^{k} \Phi_{n_{\beta}},
\label{eq: parton_wf}
\end{equation}
where $\beta$ labels the parton species. The parton state described by the wave function given in Eq.~\eqref{eq: parton_wf} is referred to by the shorthand notation ``$n_{1}n_{2}{\cdots} n_{k}$". The partons are unphysical entities so they need to be glued back together to recover the physical electrons. At the level of wave functions, this gluing procedure is implemented by setting the different parton species coordinates equal to the electron coordinate (each $\Phi_{n_{\beta}}$ in Eq.~\eqref{eq: parton_wf} is built up of all the electronic coordinates.). All the partons are exposed to the same external magnetic field as the underlying electrons and each parton species has the same density as that of the electrons. Therefore, the charge of the $\beta$ parton $q_{\beta}{=}{-}e\nu/n_{\beta}$. The constraint that the sum of the parton charges should add to that of the electron relates the parton fillings to the electronic filling, i.e., $\sum_{\beta{=}1}^{k}q_{\beta}{=}{-}e{\implies}\nu{=}[\sum_{\beta{=}1}^{k}n_{\beta}^{{-}1}]^{{-}1}$. 

Looking back at the wave functions in Eqs.~\eqref{eq: Laughlin_gs} and \eqref{eq: Jain_wf_gs}, it follows that all of the Abelian Laughlin and Jain states are parton states. The parton theory can also describe non-Abelian states. In general, a parton state with a repeated factor of $n$, with $|n|{\geq}2$ hosts non-Abelian excitations~\cite{Wen91}. The simplest non-Abelian parton state -- the $221$ state~\cite{Jain89b} at $\nu{=}1/2$ -- can be interpreted as an $f$-wave superconductor of CFs~\cite{Faugno19}. Recent identification of a parton state that lies in the same universality class as the aPf state but provides a better microscopic representation of the exact Coulomb ground state at $5/2$~\cite{Balram18} has contributed to the resurgence of interest in partons. Furthermore, it has been shown that certain high-energy excitations of the states in the Jain sequence lie beyond the scope of the CF theory~\cite{Nguyen21a} while they lend themselves to a description in terms of partons~\cite{Balram21d}. At this stage, the parton theory promises to provide a unified framework for describing all of the quantum Hall effects.

\subsection{Effective field theory}\label{sec:field}

As we mentioned in Sec.~\ref{sec: incompressible_fluids}, many of the exotic properties of FQH states originate from the emergent Chern-Simons gauge field.  Based on this idea, a field-theoretic approach to FQH systems was pioneered by Zhang, Hansson and Kivelson~\cite{Zhang89} who showed, using a Cherns-Simons field theory, that the FQH effect at Laughlin fractions can be mapped onto a superfluid of composite bosons. Alternatively, the Chern-Simons field theory of CFs was introduced by Lopez and Fradkin~\cite{Lopez91} and further developed by Halperin, Lee and Read~\cite{Halperin93} and extended via a Hamiltonian formalism by Murthy and Shankar~\cite{Murthy03}. Detailed treatment of these theories is  beyond the scope of the current chapter. 
Below we discuss two important recent developments in formulating an effective field theory of FQH systems: the role of PH symmetry and geometrical degrees of freedom.  

The interest in the field-theoretic description of FQH systems, in particular in the role played by PH symmetry, has recently been renewed by geometric resonance experiments, which probe the radius of the cyclotron orbit of CFs. A series of measurements~\cite{Kamburov14b, Hossain20} have shown that the CF Fermi wave vector $k^{*}_{F}$ is determined by the density of minority carriers in the LLL, i.e., the CF Fermi wave vector is set by the electron density for $\nu{\leq}1/2$ and by the hole density for $\nu{>}1/2$. In other words, the quantity $k^{*}_{F}\ell$ is PH-symmetric. The state that underpins the HLR theory does not reside entirely in the LLL and thus it does not obey PH symmetry manifestly. This puzzle is resolved by going beyond the mean-field level, where it has been shown that the HLR theory produces results that are PH-symmetric and consistent with experiment~\cite{Wang17}. 

In parallel, Son~\cite{Son15} proposed a field theory that is valid in the vicinity of half-filling and has a built-in PH symmetry. In Son's theory, CFs are Dirac particles whose density is controlled by the external magnetic field as opposed to by the density of electrons. The PH symmetry of electrons acts as a time-reversal symmetry on Dirac CFs. This results in an absence of $2k_{F}$ back-scattering that was subsequently observed in DMRG simulations~\cite{Geraedts17}, providing evidence in favor of Son's theory. One can ask how different are the predictions of Son's theory from that of the HLR theory? Surprisingly, it turns out that the refined version of HLR theory is consistent with Son's theory on many aspects~\cite{Wang17}.  

Interestingly, Son's construction suggests a PH-symmetric gapped state at half-filling -- dubbed the ``PH symmetric Pfaffian'' (PH-Pf). The PH-Pf state is topologically distinct from both the Moore-Read state and aPf state. Intriguingly, recent thermal Hall~\cite{Banerjee18} and heat transport measurements~\cite{Dutta21} at $\nu{=}5/2$ appear to be consistent with only the PH-Pf order. Two candidate wave functions for the PH-Pfaffian phase have been considered in the literature~\cite{Jolicoeur07, Zucker16}, both with a high degree of PH symmetry. The two wave functions also have high overlaps with each other for small systems indicating that they likely describe the same phase. However, numerical studies find no evidence for the existence of the PH-Pf phase in clean systems~\cite{Balram18, Mishmash18}. Understanding the heat transport measurements at $\nu{=}5/2$ in light of the numerical results presents one of the foremost challenges in the field.

Apart from PH symmetry, it has recently become apparent that any  effective field theory capable of capturing dynamical response of FQH phases must include the \emph{geometric} degrees of freedom  FQH systems, which are beyond the remit of purely topological field theories.  An important example of a geometric property is the Wen-Zee shift~\cite{Wen92},  related to the response function known as Hall viscosity~\cite{Avron95, Read09}. This has highlighted the need for a deeper understanding of ``quantum geometry" in FQH fluids~\cite{Haldane11}. The quantum geometry is associated with the magnetoroton (GMP) collective excitation of FQH fluids in the long-wavelength limit (cf. Sec.~\ref{sec:magnetoroton}). In this limit, the GMP mode can be interpreted as a quadrupolar density-wave deformation of the ground state and it has been incorporated into a gravitational field theory in the form of a spin-2 field coupled to an ambient geometry~\cite{GromovSon}. Occasionally, this emergent spin-2 degree of freedom is also referred to as ``FQH graviton"~\cite{Yang12b, Golkar16,Liou2019}. 
The dynamics of the spin-2 modes~\cite{Liu2018} can be induced by mass anisotropy~\cite{Yang12c,XinWanPhysRevB.86.035122,Ghazaryan15} or by tilting the magnetic field~\cite{Chakraborty1989, Papic13}. In gapless CFL states, the geometric effects manifest through the shape of the Fermi contour~\cite{Gokmen10, Jo17}. The investigation of geometric responses of FQH systems to the variations of ambient geometry represents an active research direction~\cite{you2014theory, BradlynRead, CanLaskinWiegmann}.

\subsection{Entanglement-based approaches}
\label{sec: entanglement}

Quantum entanglement has recently emerged as a powerful probe of topological order.  One characterizes entanglement by evaluating the system's reduced density matrix, obtained by partitioning the FQH system into two halves and tracing out one of the subsystems. As first shown by Li and Haldane~\cite{LiHaldane}, the spectrum of the reduced density matrix -- the ``entanglement spectrum" -- contains information about the edge excitations of the system and it can be used to infer the CFT that governs the FQH state. 

For FQH wave functions such as the Laughlin or Moore-Read, the number of levels in the entanglement spectrum as a function of  momentum in the subsystem becomes universal as the subsystem size increases and it corresponds to the edge spectrum of the corresponding CFT.  This ``bulk-boundary" correspondence suggests that the bulk  of the system and the physical edge are related to each other~\cite{Dubail12a}. On the other hand, the entanglement spectrum of a physical ground state of the Coulomb interaction is found to contain both a universal CFT part and a non-universal part. The latter is separated by a ``entanglement gap"~\cite{Thomale2010}, which is argued to remain finite in a gapped FQH phase, signaling the robustness of topological order. The entanglement spectrum has found broad use as a tool for topological characterization of phases of matter beyond FQH states discussed in this chapter.

Finally, as first shown by Zaletel and Mong~\cite{Zaletel12}, certain FQH wave functions that can be expressed as correlators of a CFT can also be turned into a  matrix-product state (MPS) if one picks the basis of  LL orbitals on a cylinder geometry.
This leads to a significant advantage in extracting physical properties from FQH wave functions. Exact MPS representations have been found for many states discussed above~\cite{Estienne2013}. Beyond studies of FQH wave functions, the MPS method has been generalized into a variational density-matrix renormalization group scheme for approximating ground states of realistic systems with Coulomb interactions~\cite{Zaletel13}. 

\begin{table*}[htb]
\centering
\begin{tabular}{|c|c|c|c|c|c|}
\hline 
filling, $\nu$ & state 	& shift, $\mathcal{S}$ & chiral central charge, $c_{-}$ &	quasiparticle charge, $\mathcal{Q}_{qp}$ & GSD	\\ \hline\hline
$1/(2p+1)$    & $1^{2p+1}$, Laughlin   &   $2p+1$				&  $1$  		& $1/(2p+1)$ & $2p+1$ \\ \hline
$n/(2pn\pm 1)$    & $(\pm n)1^{2p}$, Jain   &   $(\pm n)+2p$	&  $1\pm(n-1)$  		& $1/(2pn\pm 1)$ & $2pn\pm 1$ \\ \hline
$1/(2p)$     & Moore-Read   &   $2p+1$				    &  $3/2$			& $1/(4p)$  & $6p$ \\ \hline
$1/(2p)$     & $\bar{2}^{2}1^{2p+1}$, anti-Pfaffian   &   $2p-3$				&  $-1/2$			& $1/(4p)$  & $6p$ \\ \hline 
$2/5$    & $\bar{2}^{3}1^{4}$, anti-$\mathbb{Z}_3$ Read-Rezayi &  $-2$				&  $-4/5$  & $1/5$ & $10$  \\ \hline  
\end{tabular} 
\caption{\label{tab: one_component_states_properties} Summary of topological properties of main single-component FQH states considered in this chapter. For the states captured by parton theory, we also show their parton label according to Eq.~(\ref{eq: parton_wf}). Some of the quantities in this Table can be accessed experimentally: filling factor $\nu$ is related to the electrical Hall resistance $R_{xy}$ via Eq.~(\ref{eq:Hallresistance}),  chiral central charge $c_{-}$ is related to the thermal Hall conductance $\kappa_{xy}$ via Eq.~(\ref{eq:thermalHall}) (any filled LLs provide an additional contribution), $\mathcal{Q}_{qp}$ is the charge of fundamental quasiparticle in units of the electron charge which can be accessed via shot-noise experiments or by charge-sensing using a single-electron transistor, the shift $\mathcal{S}$ on the sphere is related to the Hall viscosity $\eta_{H}$. The ground state degeneracy (GSD) is quoted for the torus. The composite fermion Fermi liquid (CFL) is not expected to carry a quantized $\kappa_{xy}$ since its bulk is gapless.
}
\end{table*}

\section{Conclusions}
\label{sec: conclusions}

In this chapter we presented an introduction to the FQH effect observed in semiconductor systems. While it was impossible to give a historically exhaustive account of the entire field, we hope that the multitude of examples illustrates the great richness of emergent many-body phenomena and the diversity of approaches that have been developed over the years in order to understand the novel physics associated with the FQH effect. To emphasize this diversity, our presentation has focused on explaining  the experimental observations through model states, which distil the essence of FQH physics into elegant wave functions that can be studied either analytically or in computer simulations. Thus, we conclude by presenting a summary of the properties of the main single-component FQH phases discussed above in Table~\ref{tab: one_component_states_properties}. 
The table lists some key topological quantum numbers that can be used to characterize an FQH phase, as discussed in Sec.~\ref{sec: incompressible_fluids}. 
Some of these quantities are accessible in experiments and others in numerics.

Throughout this chapter, we have highlighted many open problems  and research directions that remain particularly active at the time of writing (early 2022). We note in particular that the material realizations of the FQH effect continue to expand, most notably in the context of graphene and other van der Waals materials, fractional Chern insulators, as well as synthetic quantum simulators based on ultracold atoms and superconducting qubit arrays. We purposefully avoided discussing alternative material realizations of the FQH effect, as some of them are discussed in other chapters of this encyclopedia. These new material platforms may offer additional pathways to testing some of the theoretical ideas discussed above, while they will undoubtedly pose new puzzles to be understood. For a research field that is now four decades old, the FQH effect remains a remarkably active source of exciting new physics.

\section{Acknowledgements}

We would like to thank Gabor Csathy and Wei Pan for generously sharing their experimental data that we plotted in Fig.~\ref{fig:fqhetraces}. The authors express their deep sense of gratitude to all of their collaborators over the years, without whom this chapter would not have been possible. A. C. B. acknowledges the Science and Engineering Research Board (SERB) of the Department of Science and Technology (DST) for financial support through the Start-up Grant SRG/2020/000154. A. C. B. and Z. P. thank the Royal Society International Exchanges Award IES$\backslash$R2$\backslash$202052 for funding support. Some of the numerical calculations reported in this work were carried out on the Nandadevi supercomputer, which is maintained and supported by the Institute of Mathematical Science’s High-Performance Computing Center. Z. P. acknowledges support by the Leverhulme Trust Research Leadership Award RL-2019-015.

\bibliographystyle{agsm}
\bibliography{biblio_fqhe.bib}

\end{document}